\documentclass[epj]{svjour}
\usepackage{graphics}
\usepackage[pdftex]{graphicx}
\usepackage{amsmath}
\usepackage{amssymb}
\usepackage{lineno}

\begin{document}
%
\title{Observation of a structure in the $M_{p\eta}$ invariant mass distribution near 1700 MeV/c$^2$ in the $\mathbf{\gamma p \rightarrow p \pi^0 \eta} $ reaction}

\author{
V.~Metag$^{1}$,~M.~Nanova$^{1}$,~J.~Hartmann$^{2}$, ~P.~Mahlberg$^{2}$,~F.~Afzal$^{2}$,~C.~Bartels$^{2}$,~D.~Bayadilov$^{2,6}$,~R.~Beck$^{2}$,\\
~M.~Becker$^{2}$,~E.~Blanke$^{2}$,~K.-T.~Brinkmann$^{1}$,~S.~Ciupka$^{2}$,~V.~Crede$^{3}$,~M.~Dieterle$^{4}$,~H.~Dutz$^{5}$,~D.~Elsner$^{5}$,\\~F.~Frommberger$^{5}$,~A.~Gridnev$^{6}$,
~M.~Gottschall$^{2}$,~M.~Gr\"uner$^{2}$,~Ch.~Hammann$^{2}$,~J.~Hannappel$^{2}$,~W.~Hillert$^{5,a}$,~J.~Hoff$^{2}$,\\~Ph.~Hoffmeister$^{2}$,~Ch.~Honisch$^{2}$,~T.~Jude$^{5}$,~H.~Kalinowsky$^{2}$,~F.~Kalischewski$^{2}$,~I.~Keshelashvili$^{4,b}$,~B.~Ketzer$^{2}$,~P.~Klassen$^{2}$,\\~F.~Klein$^{5}$,~K.~Koop$^{2}$,~P.~Kroenert$^{2}$,~B.~Krusche$^{4}$,~M.~Lang$^{2}$,~I.~Lopatin$^{6}$,~F.~Messi$^{5}$,~W.~Meyer$^{7}$,~B.~Mitlas\'oczky$^{2}$,\\ ~J.~M\"uller$^{2}$,~J.~M\"ullers$^{2}$,~V.~Nikonov$^{6,+}$,~V.~Novinsky$^{6}$,~R.~Novotny$^{1}$,~D.~Piontek$^{2}$,~G.~Reicherz$^{7}$,~L.~Richter$^{2}$,~T.~Rostomyan$^{4}$,\\
~S.~Runkel$^{5}$,~B.~Salisbury$^{2}$,~A.~Sarantsev$^{6}$,~D.~Schaab$^{2}$,~Ch.~Schmidt$^{2}$,~H.~Schmieden$^{5}$,~J.~Schultes$^{2}$,~T.~Seifen$^{2}$,~V.~Sokhoyan$^{2,c}$,\\~C.~Sowa$^{7}$,~K.~Spieker$^{2}$,~N.~Stausberg$^{2}$,~A.~Thiel$^{2}$,~U.~Thoma$^{2}$,~T.~Triffterer$^{7}$,~M.~Urban$^{2}$,~G.~Urff$^{2}$,~H.~van~Pee$^{2}$,~M.~Wagner$^{2}$, \\
~D.~Walther$^{2}$,~Ch.~Wendel$^{2}$,~D.~Werthm\"uller$^{4,d}$,~U.~Wiedner$^{7}$,~A.~Wilson$^{2,3}$,~A.~Winnebeck$^{2}$,~L.~Witthauer$^{4}$, and Y.~Wunderlich$^{2}$\\
(The CBELSA/TAPS Collaboration)
\mail{Volker.Metag@exp2.physik.uni-giessen.de}}
\titlerunning{ $\gamma p \rightarrow p \pi^0 \eta$: structure at $M_{p\eta} \approx$ 1700 MeV/c$^2$}
\authorrunning{V. Metag \textit{et al.}}

\institute{
{$^{1}$II. Physikalisches Institut, Universit\"at Gie{\ss}en, Germany}\\
{$^{2}$Helmholtz-Institut f\"ur Strahlen- und Kernphysik, Universit\"at Bonn, Germany}\\
{$^{3}$Department of Physics, Florida State University, Tallahassee, FL, USA}\\
{$^{4}$Departement Physik, Universit\"at Basel, Switzerland}\\
{$^{5}$Physikalisches Institut, Universit\"at Bonn, Germany}\\
{$^{6}$National Research Centre "Kurchatov institute", Petersburg Nuclear Physics Institute Gatchina, Russia}\\
{$^{7}$Physikalisches Institut, Universit\"at Bochum, Germany}\\
{$^{a}$Current address: Institute of Experimental Physics, University of Hamburg, Germany}\\
{$^{b}$Current address: Institut f\"ur Kernphysik, Forschungszentrum J\"ulich, Germany}\\
{$^{c}$Current address: Institut f\"ur Kernphysik, Universit\"at Mainz}\\
{$^{d}$Current address: Paul Scherrer Institut, Villigen PSI, Switzerland}\\
{$^{+}$ deceased}
}

\date{Received: date / Revised version: date}
%
\abstract{The reaction $\gamma p \rightarrow p \pi^0 \eta$ has been studied with the CBELSA/TAPS detector at the electron stretcher accelerator ELSA in Bonn for incident photon energies from threshold up to 3.1 GeV. This paper has been motivated by the recently claimed observation of a narrow structure in the $M_{N\eta}$ invariant mass distribution at a mass of 1678 MeV/c$^2$. The existence of this structure cannot be confirmed in the present work. Instead, for $E_{\gamma}$ = 1400 - 1500 MeV and the cut $M_{p\pi^0} \le 1190 $ MeV/c$^2$ a statistically significant structure in the $M_{p\eta}$ invariant mass distribution near 1700 MeV/c$^2$ is observed with a width of $\Gamma\approx 35$ MeV/c$^2$  while the mass resolution is $\sigma_{res}$ = 5 MeV/c$^2$. Increasing the incident photon energy from 1420 to 1540 MeV this structure shifts in mass from $\approx $ 1700 MeV/c$^2$  to $\approx$ 1725 MeV/c$^2$ ; the width increases to about 50 MeV/c$^2$  and decreases thereafter. The cross section associated with this structure reaches a maximum of $\approx$ 100 nb around $E_{\gamma} \approx$ 1490 MeV (W $\approx $ 1920 MeV), which coincides with the $p a_0$ threshold. Three scenarios are discussed which might be the origin of this structure in the $M_{p\eta}$ invariant mass distribution. The most likely interpretation is that it is due to a triangular singularity in the $\gamma p \rightarrow p a_0 \rightarrow p \pi^0 \eta$ reaction. }
%
\maketitle

\section{Introduction}
\label{intro}

As for any complex system, the excitation energy spectrum of the nucleon provides information on the interaction among its constituents. The description of this excitation energy spectrum is a challenging task for Quantum Chromodynamics (QCD), the theory of the strong interaction. While at high momentum transfers ($\ge $10 GeV/c) strong interaction phenomena can be rather successfully described in perturbative treatments because of the small coupling strength $\alpha_s$, these methods fail at momentum transfers of the order of 1 GeV/c where  $\alpha_s$ approaches unity. Numerous QCD inspired model calculations and QCD-lattice calculations have been performed, but a detailed understanding of the excitation energy spectrum of the nucleon is only slowly emerging \cite{Crede_Roberts,Klempt_Richard,Edwards}. In particular, at excitation energies W of about 2 GeV where many broad resonances strongly overlap, the number of observed states is much smaller than theoretically expected, causing the problem of the so called {\it missing resonances}  \cite{Beck_Thoma}.

Experimentally, pion- and photon- induced reactions have provided a wealth of information on the excitation energy spectrum of the nucleon. In particular photon induced reactions studied at ELSA, GRAAL, JLab, LNS (Tohoku), MAMI, and Spring8 have recently extended our knowledge \cite{Ireland}. Partial wave analyses of these data have provided information on nucleon resonances, their mass, widths and decay modes. Up to an excitation energy of 3~GeV, 20~N* (I=1/2) and 12~$\Delta$(I = 3/2) quite well established resonances with 3 or 4 star ranking are currently listed by the particle data group \cite{PDG} together with 7~N* and 10~$\Delta$ resonances which are less well established. 

While low-lying excited states of the nucleon mostly decay via the emission of a single meson, preferentially either a $\pi$ or an $\eta$ meson, higher-lying states with masses in the W $\approx $ 2~GeV range tend to decay via cascades through intermediate excited states, leading to multi-meson final states. In particular reactions like $\gamma p \rightarrow p \pi^0  \pi^0$ \cite{Assafiri,Thoma_2pi,Kashevarov_2pi,Zehr,Oberle,Dieterle,Sokhoyan_2pi_PLB,Sokhoyan_2pi,Thiel} and $\gamma p \rightarrow p \pi^0  \eta$ \cite{Ajaka,Kashevarov_pi-eta,Gutz,Kaeser,Sokhoyan_pi_eta,Metag_Nanova} have been studied in detail. Being an isospin singlet, the $\eta$ meson can only be emitted in transitions between either two N* or $\Delta$ resonances, introducing additional sensitivity in studies of decay modes.

The widths of resonances are typically of the order of 100~MeV/c$^2$  or more \cite{PDG}. Thus it was quite surprising that Kuznetsov et al. \cite{Kuznetsov} reported a narrow structure ($\Gamma \approx $ 
10 MeV/c$^2$ ) in the $M_{N\eta}$ invariant mass distribution at 1678~MeV/c$^2$ , observed in the $\gamma N \rightarrow N \pi \eta$ reaction.  These authors tentatively interpreted this narrow structure as a state in the anti-decuplet of exotic particles predicted by the Chiral Soliton Model \cite{Diakonov}. Gutz et al. \cite{Gutz} did not observe a narrow structure in their study of the $\gamma p \rightarrow p \pi^0 \eta $ reaction, however, the statistics of their measurement required a rather coarse binning of 35 MeV/c$^2$  in the $M_{p\eta}$ invariant mass, precluding the possibility of observing a narrow structure with a width of $\Gamma \approx $ 
10 MeV/c$^2$. In the analysis of the $\gamma p \rightarrow p \pi^0 \eta $ reaction by Sokhoyan et al. \cite{Sokhoyan_pi_eta} no clear indication for a narrow structure in the region of $M_{p \eta}$ = 1685 MeV/c$^2$  was found.  Also a more recent analysis of the A2 data by Werthm\"uller et al. \cite{Werthmuller} did not provide convincing evidence for the existence of a narrow structure. Furthermore, Anisovich et al. \cite{Anisovich} performed a full partial wave analysis including polarisation observables measured by the A2 collaboration \cite{Witthauer} in single $\eta$ photoproduction off the neutron and find an excellent description of the data without any narrow resonance.  This calls for a new attempt to clarify the situation experimentally and to either verify or refute this observation and interpretation. Therefore events for incident photon energies of $E_{\gamma}$ = 1400 - 1600 MeV have been selected from the full $\gamma p \rightarrow p \pi^0 \eta$ data sample to find out whether such a narrow structure is observed in the present experiment.
 
The paper is structured as follows: The experimental setup and the conditions of the experiment are described in section 2. Details of the data analysis are given in section 3. In section 4 the present data are compared to previous results. The main experimental results are presented in section 5. In section 6 different scenarios for the interpretation of the present data are discussed. Concluding remarks are given in section 7.


\section{Experimental setup}
\label{expsetup}

The experiment was performed at the electron stretcher accelerator ELSA in Bonn \cite{Husmann_Schwille,Hillert}. Photons were produced by scattering electrons of 3.2~GeV off a 500-$\mu$m-thick diamond radiator and impinged on a 5-cm-long LH$_2$ target. The bremsstrahlung photons were tagged in the energy range of 0.7-3.1~GeV by detecting the scattered electrons in coincidence after deflection by a tagging magnet. The recoil protons and decay photons from $\eta$ and $\pi^0$ mesons produced by the interaction in the target were detected with the combined Crystal Barrel (CB) (1320 CsI(Tl) modules) \cite{Aker} and MiniTAPS (MT) calorimeters (216 BaF$_2$ modules) \cite{TAPS1,TAPS2}. This detector setup covered polar angles of 11$^{\circ}$-156$^{\circ}$ and 1$^{\circ}$-11$^{\circ}$, respectively, and the full azimuthal angular range, thereby covering 96$\%$ of the full solid angle. In the angular range of 11.2$^\circ$-27.5$^\circ$, the so-called Forward Plug (FP), the CB modules were read out by photomultipliers, providing energy and time information while the rest of the CB crystals were read out by photodiodes with energy information only. The photon energy resolution is $\sigma/E\approx 2.4\%$/E[GeV]$^{1/4}$ and the angular resolution is $\approx 1.8^0$. Because of the high granularity and the large solid-angle coverage the detector system was ideally suited for the detection and reconstruction of multi-photon events. 

At polar angles of $1^{\circ}$-$11^{\circ}$, protons were registered in plastic scintillators in front of the MiniTAPS forward wall. In the angular range of 11.2$^{\circ}$-27.5$^{\circ}$ charged particles were detected in plastic scintillators in front of the CB modules and for 21$^{\circ}$-167$^{\circ}$ they were identified in a three-layer scintillating fibre array \cite{Suft}. The polar angular resolution for proton detection is $\sigma =1^{\circ}$ in MiniTAPS and $\sigma \approx $ 3$^{\circ}$ at laboratory angles 11$^{\circ}$-156$^{\circ}$, given by the size of the crystals and the length of the target cell.

The photon flux through the target was determined by counting the photons reaching the Gamma Intensity Monitor (GIM) \cite{Gottschall}, a $4\times4$ matrix of PbF$_2$ crystals at the end of the setup, in coincidence with electrons registered in the tagging system. The total rates in the tagging system were $\approx$ 17~MHz. Since the GIM efficiency decreased at high rates ($\gg $1 MHz), the Flux monitor (FluMo) detector, operating at much lower rates by measuring e$^+$e$^-$ pairs from photon conversion in a Pb foil, served to monitor the rate-dependent GIM efficiency. Systematic errors associated with the photon flux determination using the GIM were estimated to be about 5-10$\%$  because of an energy dependence of the GIM efficiency. The polarisation of the incident-photon beam, obtained by using the diamond crystal as a bremsstrahlung target, was not exploited in the present analysis of the data. In the full data set the polarisation was averaged out since the polarisation was regularly switched after each run during the data taking. A CO$_2$ gas-Cherenkov detector with a refractive index of n=1.00043 was used to veto electromagnetic background (electrons and positrons). The data were collected during a data-taking period of about 1500~h corresponding to a total integrated luminosity of 20~pb$^{-1}$. 
 
 The first-level trigger required two or more hits in MiniTAPS or Forward Plug and no hit in the gas-Cherenkov detector. If there were no hits in MT or FP, events with a charged hit in the inner detector were also included in the first level trigger and further processed if additionally two energy deposits in the CB were identified by the fast cluster encoder (FACE) in the second-level trigger. Also events with one hit in MT or FP  and 1 cluster reconstructed in FACE were accepted provided there was no hit in the gas-Cerenkov detector. The dead time introduced by the gas-Cherenkov detectors was 4$\%$.  A more detailed description of the detector setup and the running conditions can be found in \cite{Gottschall,Hartmann}. 

 \section{Data analysis}
 \label{sec:ana}
 
 \subsection{Data selection}
 \label{selection}
 
 \begin{figure}
\begin{center}
 \resizebox{0.5\textwidth}{!}
 {\includegraphics[width=5.0cm]{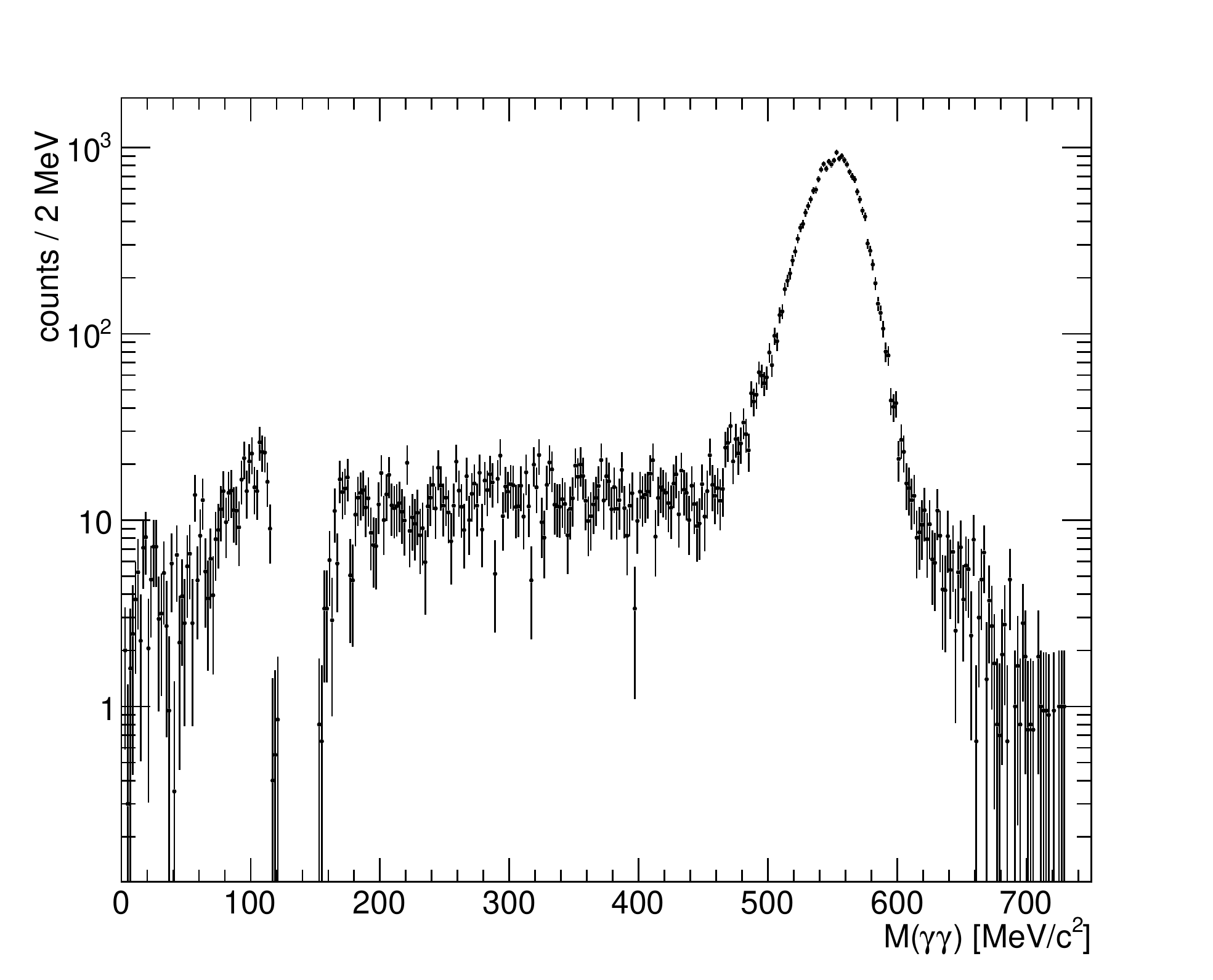}}
 \caption{Invariant mass spectrum $M_{\gamma \gamma}$ for confidence level CL($\gamma p \rightarrow p_{miss} \pi^0 \gamma \gamma) \ge $ 0.2 and CL($\gamma p \rightarrow p_{miss} \pi^0 \pi^0) \le$ 0.01. The hole in the spectrum near the $\pi^0$ mass of 135 MeV/c$^2$ demonstrates that $\pi^0 \pi^0$ events are effectively removed from the data sample. The plot is for the incident photon energy range 1400 - 1600 MeV.}
\label{fig:Mgg}
\end{center}
\end{figure}

\begin{figure*}
\begin{center}
 \resizebox{1.0\textwidth}{!}
 {\includegraphics[width=10.0cm,clip]{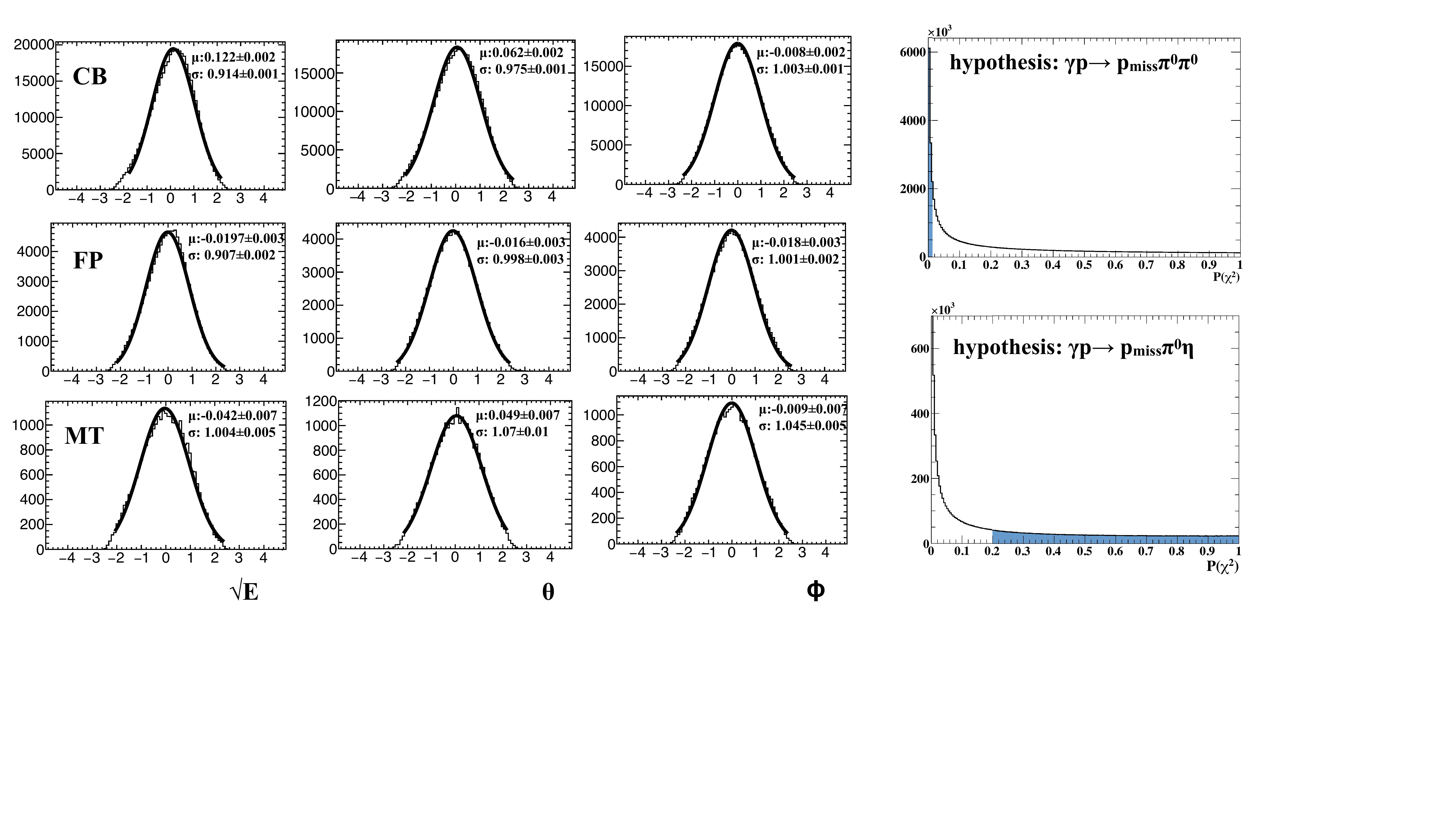}}
 \vspace{-2.5cm}
 \caption{Left: pull distributions from the $\gamma p \rightarrow p_{miss} \pi^0 \eta$ fit to the data in the incident photon energy range $E_{\gamma}$= 1400 - 1600 MeV. Top row: particles in the Crystal Barrel; middle row: particles in the Forward Plug; bottom row: particles in MiniTAPS. Left to right: pulls in square root of energy, $\theta, \phi$. The pulls are displayed for events with a confidence level P$(\chi^2) \ge $ 0.1. Right: confidence level distributions for the hypotheses $\gamma p \rightarrow p_{miss} \pi^0 \pi^0$ and $\gamma p \rightarrow p_{miss} \pi^0 \eta$, imposing energy and momentum conservation. The final state proton is treated as a missing particle. Events with probabilities in the blue shaded areas (CL$(p \pi^0 \eta ) \ge$ 0.2 and CL$(p \pi^0 \pi^0) \le$ 0.01) are retained for further analysis.}
\label{fig:kinfit}
\end{center}
\end{figure*}

In the offline analysis, events from the $\gamma p \rightarrow p \pi^0 \eta $ reaction were preselected via the two photon decays of the $\pi^0$ and $\eta$ mesons with branching ratios of 98.8$\%$ and 39.4$\%$, respectively \cite{PDG}; thus events with one charged hit and four neutral clusters were processed further. Cluster energies were determined by summing the energy deposits in contiguous calorimeter modules. 
Energy deposits from split-offs were suppressed by requesting a minimum cluster energy of 25 MeV in MT and 20 MeV in CB, respectively. There were two subclasses of events: those with energy deposition of protons in the charged particle detectors and the crystals and those with proton hits only in the charged particle detectors. Combining both event classes leads to a flat detector acceptance $\star$ efficiency and an almost complete coverage of phase space, as discussed in section \ref{acc_eff}. Knowing the initial state of the reaction (photon of known energy in the beam (z)-direction and target proton at rest), energy and momentum balance permitted the treatment of the proton as missing particle and to calculate its 4-momentum vector. Only events were retained where the calculated mass of the proton was within 850 - 1030 MeV/c$^2$ . 
 
 A time coincidence between the registered final state particles and the scattered electron in the tagger was required. Depending on the detector components the time resolution varied between $\sigma$ = 0.27 ns to 1.9 ns \cite{Gottschall}. Random background events were removed by side-band subtraction. The peak-to-total ratio in the prompt peak was 13$\%$.
 
\subsection{Kinematic fit}
\label{kin_fit}
\begin{figure}
\begin{center}
 \resizebox{0.5\textwidth}{!}
 {\includegraphics[width=6.0cm]{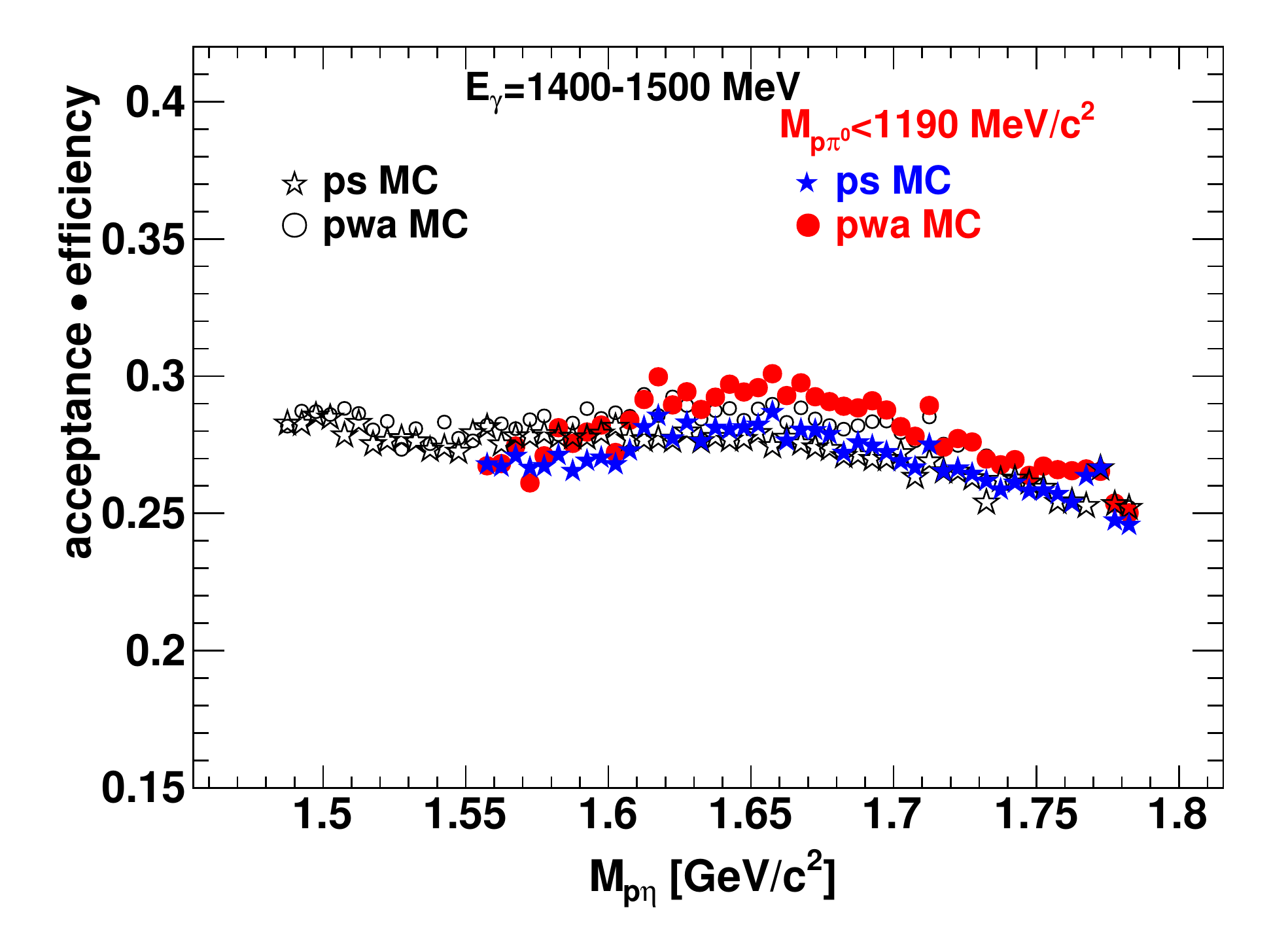}}
 \caption{Acceptance $\star$ efficiency for events of the  $\gamma p \rightarrow  p  \pi^0 \eta$ reaction as a function of the invariant mass $M_{p\eta}$ obtained in GEANT3 \cite{GEANT} simulations, assuming a distribution of final state particle 4-momenta according to a phase space distribution (stars) and to the predictions of a partial wave analysis (PWA) \cite{Klempt}(circles). The full symbols represent the probability to reconstruct  $\gamma p \rightarrow p \pi^0 \eta$ events under the condition $M_{p\pi^0} \le$1190 MeV/c$^2$ , while the open symbols refer to the acceptance $\star$ efficiency without this cut. Note the suppressed zero.}
\label{fig:acc}
\end{center}
\end{figure}

Events with the $ p4\gamma$ final state mostly come from the dominant $\gamma p \rightarrow  p  \pi^0 \pi^0$ reaction and not from the reaction of interest: $\gamma p \rightarrow  p  \pi^0 \eta$. To separate events from these two reactions both these hypotheses as well as the hypothesis $\gamma p \rightarrow  p  \pi^0 \gamma \gamma$ were tested via kinematic fits over the full incident photon energy range, imposing energy and momentum conservation as well as the masses of the final state particles as constraints. Hereby, the proton was treated as a missing particle, however, the proton polar and azimuthal angle derived from the kinematic fit was requested to agree within $\pm$~8$^{\circ}$ with the polar angle and within $\pm$~9$^{\circ}$ with the azimuthal angle of the experimentally registered charged hit, respectively. Details of the kinematic fit procedure are given in van Pee et al. \cite{Pee} and Gutz et al. \cite{Gutz}. 

A kinematic fit testing the reaction channel  $\gamma p \rightarrow  p  \pi^0 \gamma \gamma$ was performed to check whether an anti-cut on the confidence level CL($\pi^0 \pi^0)$ can sufficiently suppress events from the $\gamma p \rightarrow  p  \pi^0 \pi^0$ reaction. This is demonstrated in Fig.~\ref{fig:Mgg} which shows the $\gamma \gamma$ invariant mass distribution of the events after applying the confidence level cuts CL($\pi^0 \gamma \gamma) \ge 0.2$  and CL($\pi^0 \pi^0) \le 0.01$. If all $\gamma p \rightarrow  p  \pi^0 \pi^0$ events are removed from the data sample by these confidence level cuts the $\gamma \gamma$  invariant mass distribution of $\gamma p \rightarrow  p  \pi^0 \gamma \gamma$ events should not show any counts near the $\pi^0$ mass but a clear peak at the $\eta$ mass of 547.9 MeV/c$^2$, as observed. Thus a confidence level cut CL($\pi^0 \pi^0) \le 0.01 $ appears to sufficiently suppress the $\gamma p \rightarrow  p  \pi^0 \pi^0$ background. The background-to signal ratio is 1$\%$ in a $\pm 2.5 $ $\sigma$ interval around the $\eta$ peak. The effects of $ p \pi^0 \pi^0 $ impurities in the data sample have been discussed in detail in \cite{Metag_Nanova}.
 \begin{figure*}
  \resizebox{1.05\textwidth}{!}
{\includegraphics[width=10 cm]{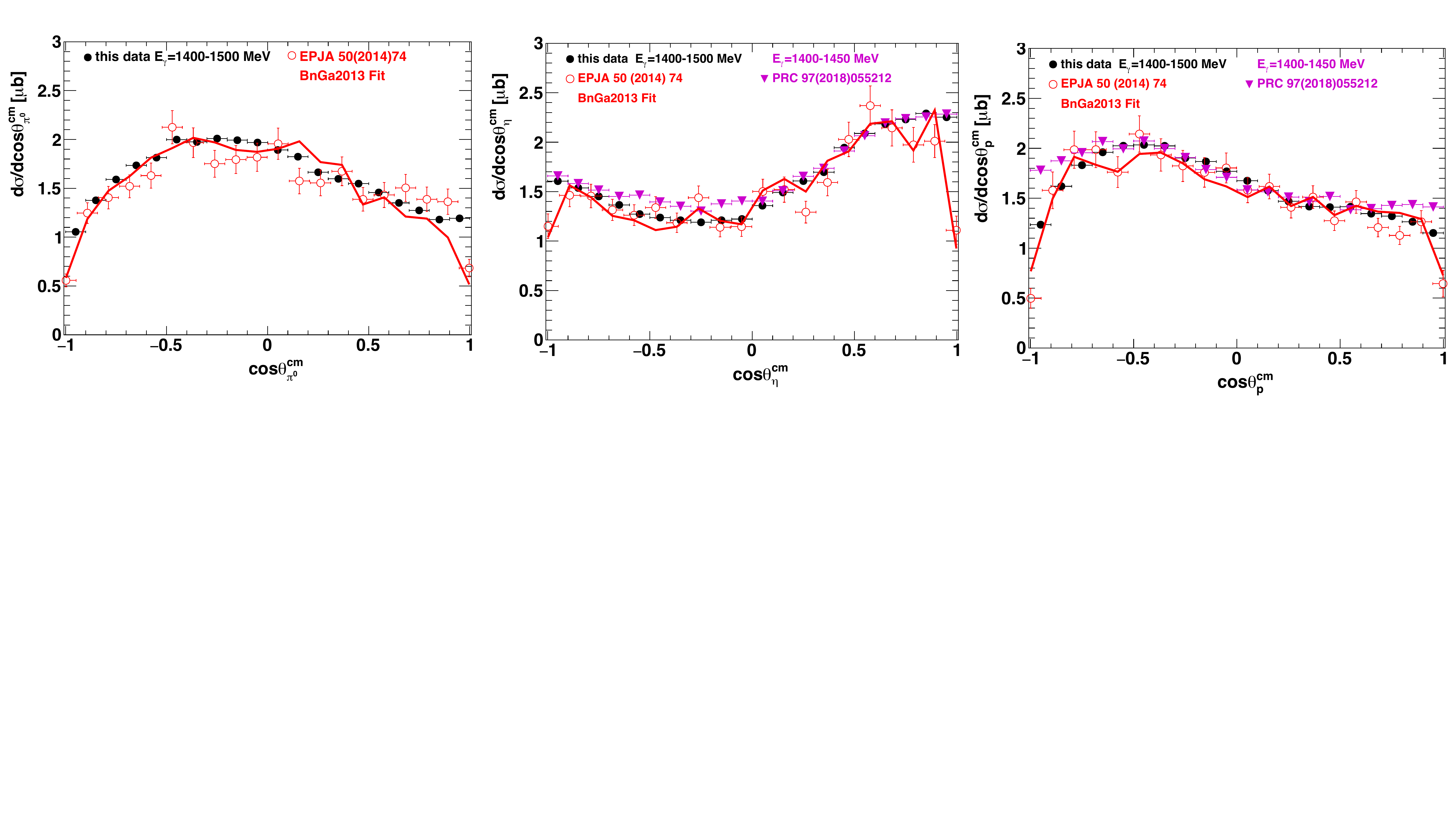}}
\vspace{-5.8cm}
 \caption{Comparison of the differential cross section for $\pi^0, \eta$ mesons and protons in the $\gamma$p centre-of-mass system for the incident photon energy range of $E_{\gamma} = 1400 - 1500$ MeV (full black points) with previous CBELSA/TAPS data \cite{Gutz} (open red points) and the PWA (red curve) \cite{Klempt}. In addition, the corresponding data of the A2 collaboration (inverse violet triangles) \cite{Sokhoyan_pi_eta} for the incident photon energy range 1400 - 1450 MeV are shown.}
\label{fig:comp_PWA_Gutz_A2}
\end{figure*}

\begin{figure*}
\begin{center}
 \resizebox{0.9\textwidth}{!}
{\includegraphics[width=6cm]{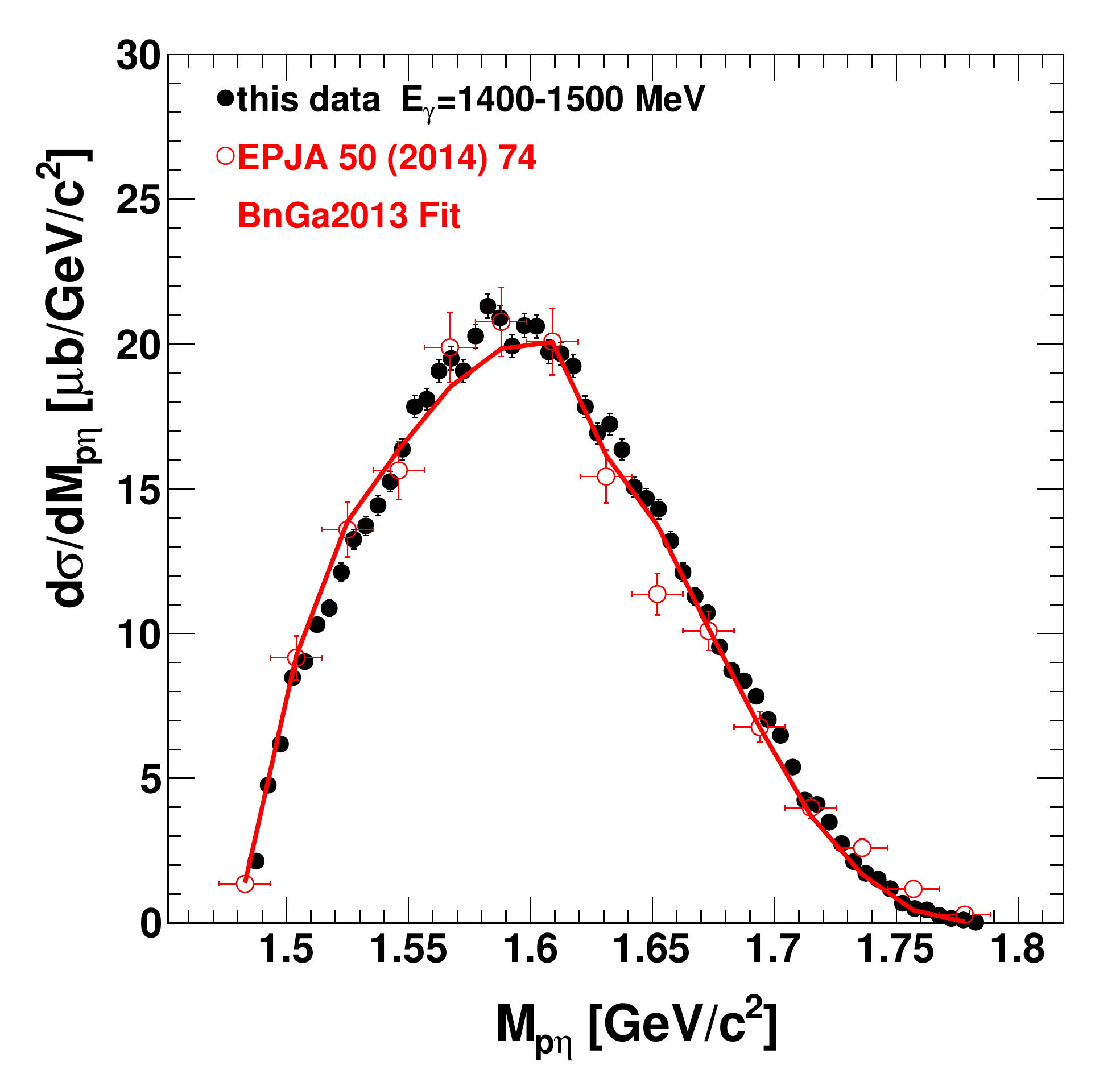}\includegraphics[width= 6cm]{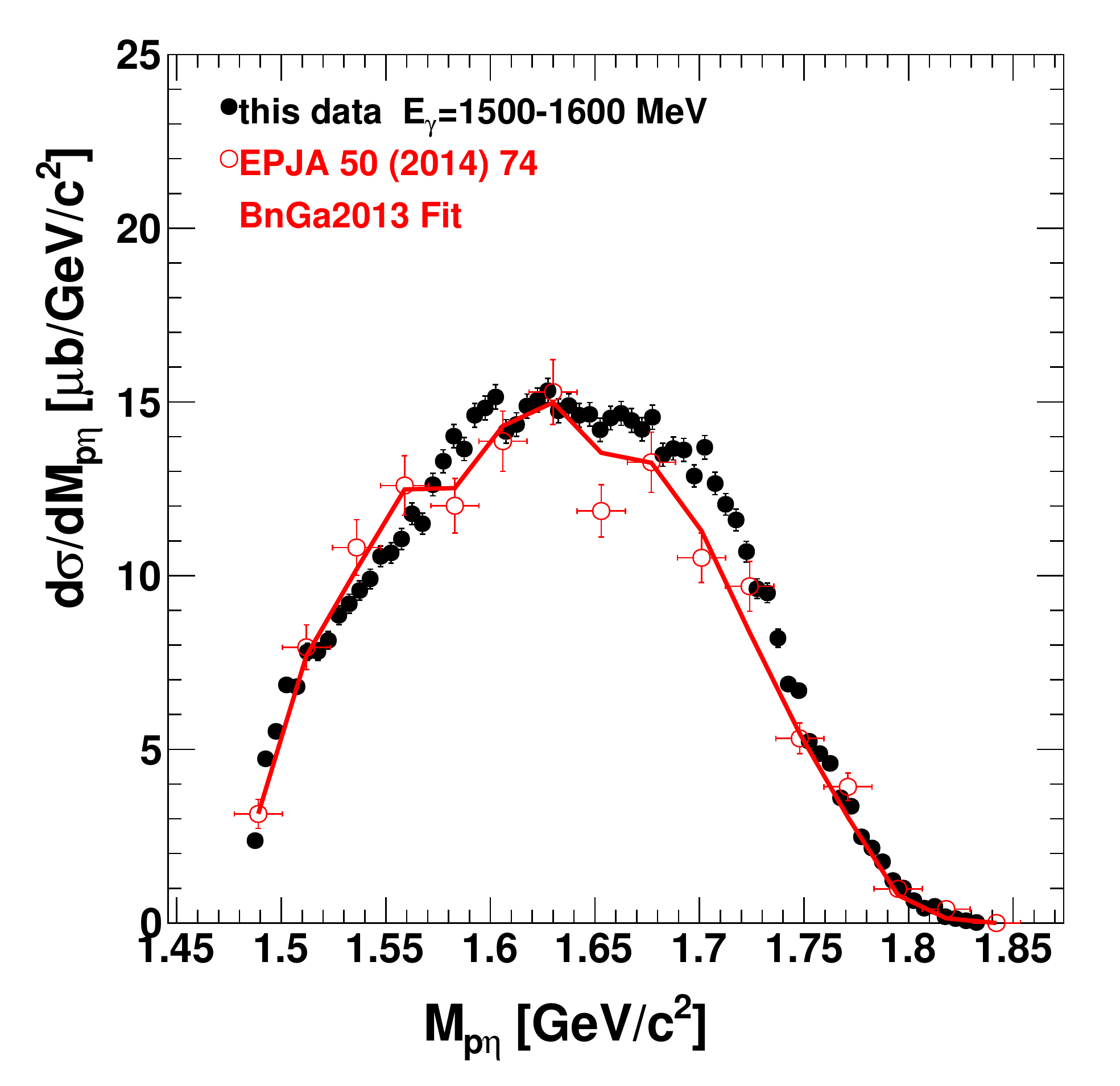}}
 \caption{$M_{p \eta}$ invariant mass distributions for $E_{\gamma}$ = 1400 - 1500 MeV and 1500 - 1600 MeV, respectively, in comparison with earlier CBELSA/TAPS data (open red points) and the Bonn-Gatchina partial wave analysis \cite{Gutz}, showing good agreement with increased statistical precision.}
\label{fig:comp_minv}
\end{center}
\end{figure*}

The quality of the kinematic fit of the $\gamma p \rightarrow p \pi^0 \eta$ reaction is illustrated in Fig.~\ref{fig:kinfit}. The pull distributions, testing the quality of the error estimation in the fit, are shown in Fig.~\ref{fig:kinfit} (left) separately for particles detected in the CB, the Forward Plug and TAPS for the square root of their energy and their polar and azimuthal angle, respectively. The fits are compatible with the expected Gaussian shape with mean $\mu$ = 0 and $\sigma $ = 1.0. The right panel of Fig.~\ref{fig:kinfit}  shows the confidence level distributions for the hypotheses $\gamma p \rightarrow p \pi^0 \pi^0$ and $\gamma p \rightarrow p \pi^0 \eta$. For the subsequent analysis events with confidence levels CL$(p \pi^0 \eta ) \ge$ 0.2 and CL$(p \pi^0 \pi^0) \le$ 0.01 are selected. The confidence level cuts represent a compromise between statistical significance and signal-to-background ratio. For the incident photon energy range of $E_{\gamma}$ = 1400 - 1600 MeV a data sample of 165 000 $\gamma p \rightarrow p \pi^0 \eta$ events has been obtained in this way.


\subsection{Correction for detector acceptance $\star$ efficiency}
\label{acc_eff}

\begin{figure*}
\begin{center}
 \resizebox{0.9\textwidth}{!}
 {\includegraphics[width=9cm]{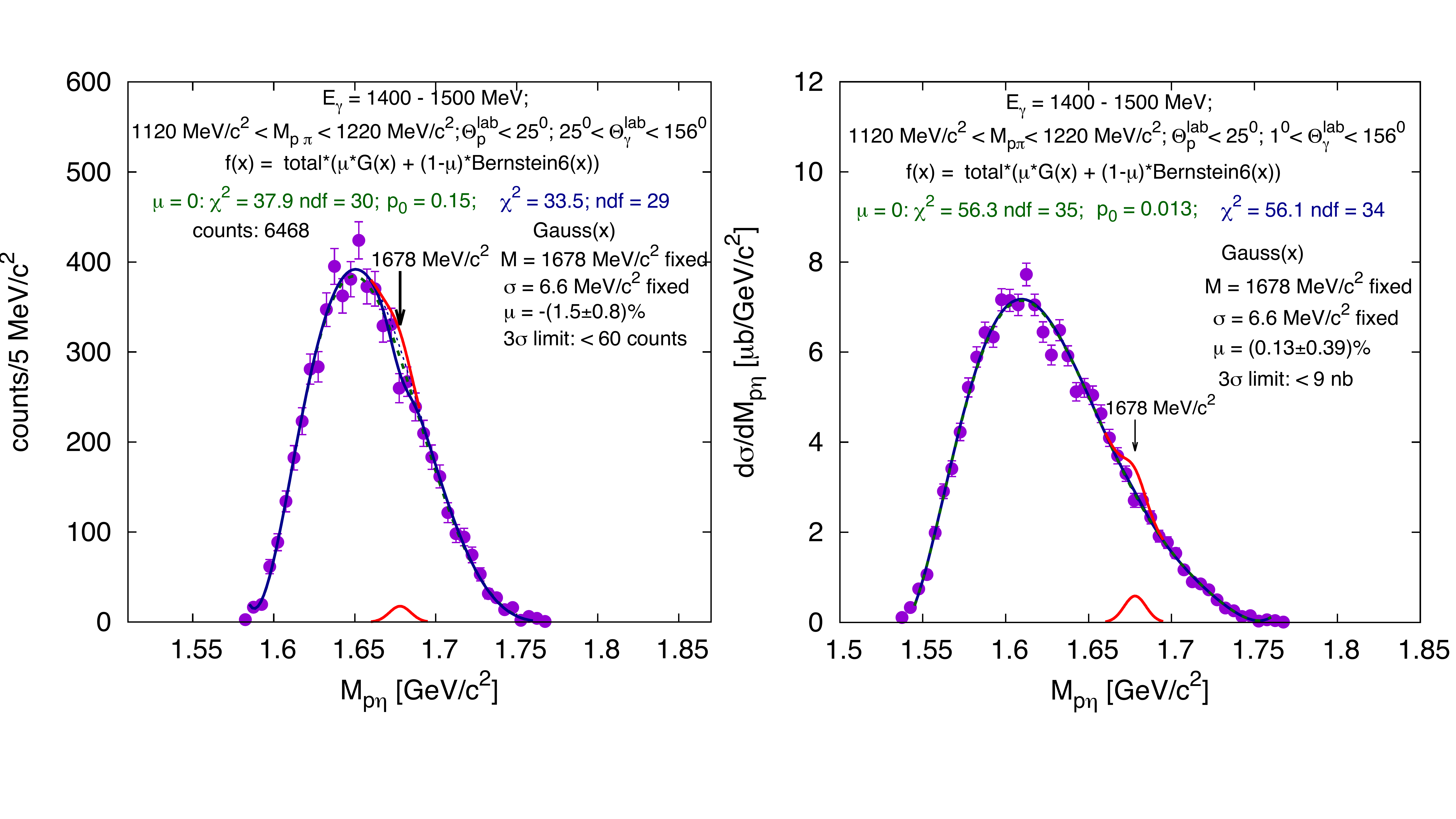}}
 \vspace{-0.8cm}
 \caption{Left: $M_{p\eta}$ invariant mass distribution under the conditions: $E_{\gamma}$ = 1400 - 1500 MeV; 1120 MeV/c$^2 \le M_{p\pi^0} \le 1220$ MeV/c$^2$ ; $\theta_p^{lab} \le 25^{\circ};  25^{\circ} \le \theta_\gamma^{lab} \le 155^{\circ}$, as applied by Kuznetsov et al. \cite{Kuznetsov}. The blue curve represents a fit of the data using a Bernstein polynomial of 6th order and a Gaussian with position and width fixed at the values reported in \cite{Kuznetsov} while the amplitude of the Gaussian is a free fit parameter. The position of the invariant mass peak reported in \cite{Kuznetsov} is marked by an arrow. In the present work an upper limit of 60 counts is deduced at the 95$\%$ confidence level for a structure at $M_{p\eta}$ = 1678 MeV/c$^2$ . Right: $M_{p\eta}$ invariant mass distribution under the conditions: $E_{\gamma} $= 1400 - 1500 MeV; 1120 MeV/c$^2 \le M_{p\pi^0} \le 1220$ MeV/c$^2$ ; $\theta_p^{lab} \le 25^{\circ};  1^{\circ} \le \theta_\gamma^{lab} \le 155^{\circ}$. The position of the invariant mass peak reported in \cite{Kuznetsov} is again marked by an arrow. An upper limit of 9 nb is deduced at a confidence level of 95$\%$. The lower red curves in both plots indicate the signals that are rejected at the 3 $\sigma$ level. Summing these signals and the respective blue fit curves gives the upper red curves. The green dashed curves (only visible near the marked position of the resonance structure in the left figure) represent fits to the data assuming a null-hypothesis, i.e. no signal  and thus a fit only with the Bernstein polynomial.}
\label{fig:anti-K}
\end{center}
\end{figure*}

All spectra shown in this paper have been corrected for the acceptance $\star$ efficiency of the CBELSA/TAPS detector system, except for Fig.~\ref{fig:anti-K} (left). The acceptance $\star$ efficiency have been determined with the CBELSA/TAPS analysis package using GEANT3 Monte Carlo simulations \cite{GEANT} with a full implementation of the detector system, including e.g., detector thresholds and trigger conditions. The $\gamma p \rightarrow p \pi^0 \eta$ reaction has been simulated, assuming a phase space distribution of the final state particles and alternatively a distribution provided by a partial wave analysis (PWA) \cite{Klempt} based on the experimental results of \cite{Gutz,Sokhoyan_pi_eta}. The detector acceptance $\star$ efficiency is given by the ratio of the number of reconstructed to generated events for each $M_{p\eta}$ invariant mass bin. The resulting  acceptance $\star$ efficiency distributions are shown in Fig.~\ref{fig:acc} with and without a cut on the $M_{p\pi^0}$ invariant mass distribution, to be discussed later. The acceptances and efficiencies obtained in the phase space and PWA simulations show a flat and smooth dependence on the $M_{p\eta}$ mass and agree within 6~$\%$, used as estimate of the systematic error for the reconstruction efficiency. In the subsequent analysis the PWA-based acceptance $\star$ efficiency has been used. The invariant mass resolution has also been determined by simulations and is found to be 5 MeV/c$^2$($\sigma$).

As an independent test of the analysis procedure, the Monte Carlo simulations and the implementation of the detector geometry, events of the simultaneously measured reaction $\gamma p \rightarrow p \eta \rightarrow p~ 6 \gamma$ have been analysed. The differential cross sections in the energy range of $E_{\gamma}$ =~1450~-~1500 MeV reported in \cite{Kashevarov_eta-eta',Crede_eta-eta'} have been reproduced within 5$\%$, demonstrating the reliability of the data analysis.

An estimate of the systematic errors is given in table \ref{tab:syst}. The systematic error of the cross sections is about $10\%$ determined by uncertainties in the reconstruction efficiency and the photon flux. The systematic error of the results presented in section \ref{structure} are about 15$\%$ estimated from variations in the fits of $M_{p\eta}$ distributions using different fit functions.

\begin{table}[h!]
\centering
\caption{Sources of systematic errors.}
\begin{footnotesize}
\begin{tabular}{cc}

reconstruction efficiency & $\approx$ 6$\%$\\
photon flux & 5-10$\%$\\
\hline
systematic error of cross sections & $\approx$ 10$\%$\\

systematic error of fits & $\approx$ 15$\%$\\

\end{tabular}
\end{footnotesize}
\label{tab:syst}
\end{table}

\section{Results}
\label{results}
\subsection{Comparison to previous results}
\label{comparison}
As a first step, the angle differential cross sections for the final state particles and $M_{p\eta}$ invariant mass distributions are compared to previous analyses of the $\gamma p \rightarrow p \pi^0 \eta$ reaction in the relevant energy regime. Fig.~\ref{fig:comp_PWA_Gutz_A2} shows a comparison of the acceptance $\star$ efficiency corrected proton, $\pi^0$ and $\eta$ angular distributions, obtained from the present data, with the ones published in \cite{Gutz} and the results of the event-based partial wave analysis (PWA) by the Bonn-Gatchina group \cite{Klempt}, based on the results of Gutz et al. \cite{Gutz}. In addition, the results of the A2 collaboration \cite{Sokhoyan_pi_eta} for incident photon energies of $E_{\gamma} $= 1400 - 1450 MeV are shown for comparison. The experimental angular distributions agree within the statistical and systematic uncertainties and are very well reproduced by the PWA-results. Integrating the differential cross sections, the total cross section of the $\gamma p \rightarrow p \pi^0 \eta$ reaction is found to be 3.23 $\mu$b and 3.16 $\mu$b for incident photon energies of 1400 - 1500 MeV and 1500 - 1600 MeV, respectively, in excellent agreement with the cross sections reported by the A2 collaboration \cite{Sokhoyan_pi_eta} and Gutz et al. \cite{Gutz}. The statistical errors are negligible and the systematic error is about 10$\%$ (s. table \ref{tab:syst}).

As shown in Fig.~\ref{fig:comp_minv}, the $M_{p\eta}$ invariant mass distributions, which play a central role in the present analysis, are also consistent with the corresponding distributions reported in \cite{Gutz}.


 \begin{figure*}
\begin{center}
 \resizebox{0.8\textwidth}{!}
 {\includegraphics[width=10cm]{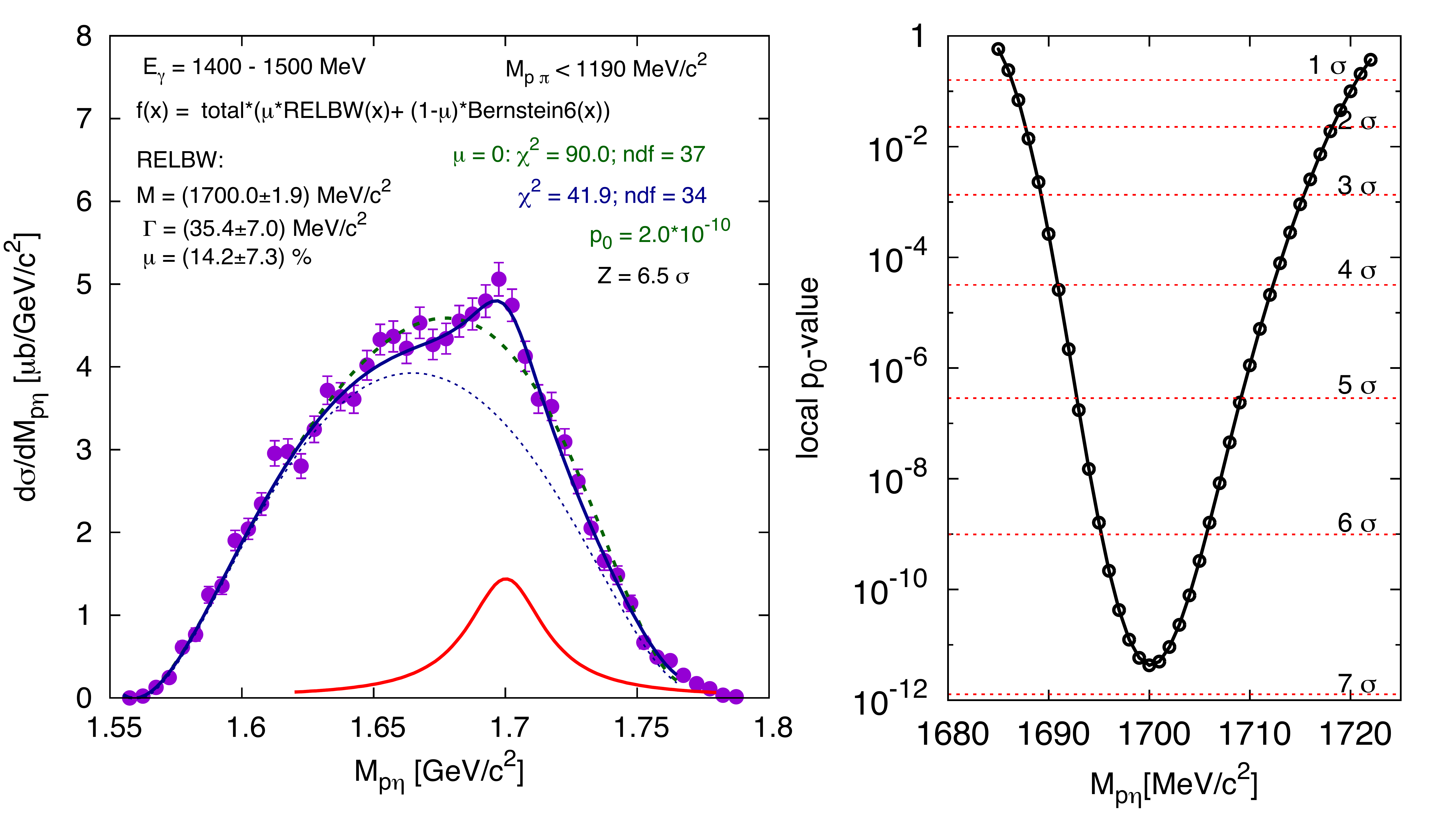}}
 \caption{Left: $M_{p\eta}$ invariant mass distribution for $E_{\gamma}$ = 1400 - 1500 MeV and $M_{p\pi^0} \le 1190 $ MeV/c$^2$ , allowing for all proton and photon laboratory angles. The blue curve represents a fit to the data using a Bernstein polynomial of 6th order to describe the physics background of decay cascades via the $\Delta$(1232) resonance and a relativistic Breit-Wigner shape for the signal. A structure is observed at $M_{p\eta} = (1700 \pm 2)$ MeV/c$^2$  with a width of $\Gamma = (35 \pm 7)$ MeV/c$^2$ . The strength of the signal corresponds to (14 $\pm 7)\%$ of the total spectrum. The significance of the signal is  6.5 $\sigma$. The dashed-blue curve represents the fitted background contribution. The dashed green curve shows the fit with the null-hypothesis (no-signal) with a probability $p_0 = 2 \cdot 10^{-10}$ (Eq. \ref{eq:p_0 value}). Right: local $p_0$ value at a given $M_{p\eta} $ mass, varied in steps of 1 MeV/c$^2$ .}
 \vspace{-0.5cm}
\label{fig:structure}
\end{center}
\end{figure*}

\begin{figure*}
\begin{center}
 \resizebox{0.9\textwidth}{!}
 {\includegraphics[,width=3.0cm]{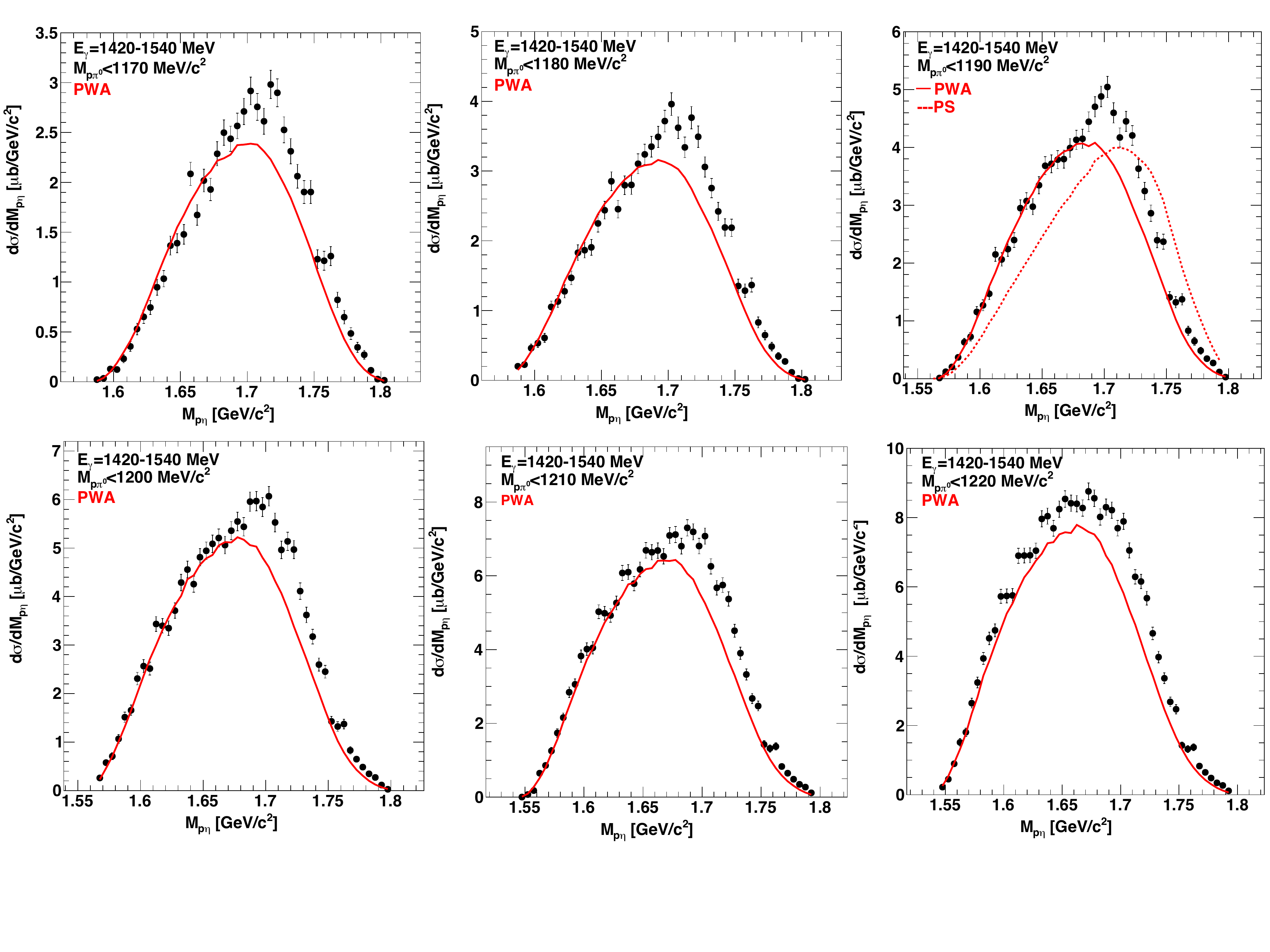}}
 \vspace{-1cm}
 \caption{$M_{p\eta}$ invariant mass distributions for the incident photon energy range $E_{\gamma}$ = 1420 - 1540 MeV and different $M_{p\pi^0}$ cuts in comparison to the corresponding distributions predicted by the partial wave analysis (PWA; red curves) \cite{Klempt}. The PWA distribution in the plot (top right) has been normalised to the experimental data in the mass range  $M_{p\eta}$ = 1570 - 1680 MeV/c$^2$. Here, a phase space distribution (dotted red curve) is shown in addition, normalised to the same area as the PWA distribution. Applying the $M_{p\pi^0}$ cuts on the PWA and phase space distributions does not produce the experimentally observed structures. The same normalisation factor has been applied to all other PWA invariant mass distributions. } 
  \label{fig:comp_PWA_cuts}
\end{center}
\end{figure*}

\subsection{Search for the reported narrow structure in the $M_{p\eta}$ invariant mass distribution at 1678 MeV/c$^2$ }
\label{anti_Kuznetsov}
Previous studies of the $\gamma p \rightarrow  p \pi^0 \eta$ reaction have established the dominance of the D$_{33}$ partial wave for incident photon energies from threshold up to about 1500 MeV \cite{Ajaka,Kashevarov_pi-eta,Kaeser_D,Doering}. Near threshold the $\Delta(1700)3/2^-$ resonance and at higher energies the $\Delta(1940)3/2^-$ resonance are predominantly populated which both decay via $\eta$ emission to the $\Delta(1232)3/2^+$ resonance with subsequent $\pi^0$ transition to the ground state. In a search for a weak and narrow structure this dominant contribution of the $\gamma p \rightarrow \Delta^* \rightarrow \eta \Delta(1232) 3/2^+ \rightarrow p \eta \pi^0 $ decay chain has to be suppressed; Kuznetsov et al. \cite{Kuznetsov} e.g. applied a cut on the $M_{N \pi} $ invariant mass distribution to reduce events from decay cascades via the $\Delta(1232)3/2^+$ resonance. 

The present data have been analysed under the identical conditions as in \cite{Kuznetsov}, i.e. the incident photon energy range is $E_{\gamma} = 1400-1500 $ MeV and the cut 1120 MeV/c$^2  \le M_{p \pi^0} \le 1220 MeV/c^2$ has been applied. The angular range for protons and photons in the laboratory is confined to  $ 1^{\circ} \le \theta_{p}^{lab} \le 25^{\circ} $ for protons and to $25 ^{\circ}\le \theta_{\gamma}^{lab} \le 155^{\circ}$ for photons, as in \cite{Kuznetsov}. The resulting $M_{p\eta}$ invariant mass distribution is shown in Fig.~\ref{fig:anti-K} (left) without acceptance correction for a direct comparison with Fig.~5 in \cite{Kuznetsov}. Although the number of events is four-times higher no significant structure is observed at the mass reported by \cite{Kuznetsov}. Assuming a Gaussian shaped signal at 1678 MeV/c$^2$ with a width of $\sigma = 4.3 $ MeV/c$^2$ ($\Gamma = 10 $ MeV/c$^2$), convoluted with the mass resolution of 5 MeV, the fit returns an intensity of $- (1.1\pm 0.6) \%$ of the total spectrum, corresponding to less than 60 counts at the 3 $\sigma$ confidence level. Allowing for all photon laboratory angles in the CBELSA/TAPS set up ($1 ^{\circ}\le \theta_{\gamma}^{lab} \le 156^{\circ}$), but maintaining all other cuts, the differential cross section $d\sigma/d$$M_{p\eta}$  shown in Fig.~\ref{fig:anti-K} right is obtained after correcting for the detector acceptance $\star$ efficiency and normalising to the photon flux. The analysis of this spectrum provides an upper limit for the cross section of a structure at 1678 MeV/c$^2$ of $<$ 9 nb at the 95$\%$ confidence level.

 \section{Observation of a structure near 1700~MeV/c$^2$ in the $M_{p\eta}$ invariant mass distribution}
\label{structure}
\subsection{$E_{\gamma} = $ 1400 - 1500 MeV}
Allowing for all proton laboratory angles and applying the cut $M_{p\pi^0} < 1190 $ MeV/c$^2$ to suppress decay cascades via the $\Delta(1232) 3/2^+ $ resonance, the $M_{p\eta}$ invariant mass spectrum shown in Fig.~\ref{fig:structure} left has been obtained for the incident photon energy range $E_{\gamma} $ = 1400 - 1500 MeV. The bulk part of the spectrum is still due to events from the dominant decay chain $\gamma p \rightarrow \Delta(1232) \eta \rightarrow p \pi^0 \eta$ which have not been fully suppressed by the $M_{p\pi^0} < 1190 $ MeV/c$^2$ cut. The spectrum exhibits a structure at  $\approx$ 1700 MeV/c$^2$. The choice of the cut at $M_{p\pi^0}$ = 1190 MeV/c$^2$ is a compromise between obtaining sufficient statistics on the one hand and a reasonable signal-to-noise ratio on the other hand, needed for more detailed studies of the structure. A fit using a Bernstein polynomial of 6th order to describe the physics background of decay chains via the $\Delta(1232)$ resonance and assuming the shape of a relativistic Breit-Wigner function with constant width for the signal gives a peak position of $M_{p\eta} = (1700 \pm 2)$ MeV/c$^2$ and a width of $\Gamma = (35 \pm 7) $ MeV/c$^2$ with an intensity of $(14 \pm 7)\%$  of the total spectrum. 

The significance of the structure is determined by evaluating the likelihood ratio $\Lambda =  L_{\text{bg}}/L_{\text{bg+s}}$, whereby $L_{\text{bg+s}}$ is the likelihood of the model assuming that the data can be described by a signal+background, while $L_{\text{bg}} $ is the likelihood of the null-hypothesis, assuming only background. Each of the two models is separately fitted to the data. According to Wilks' theorem the difference in log-likelihood for both scenarios is given by the difference in the corresponding $\chi^2$ values.

\begin{equation}
-2 \ln\Lambda =-2 (\ln L_{\text{bg+s}} -\ln L_{\text{bg}}) = \chi_{\text{bg}}^2-\chi_{\text{bg+s}}^2   \label{eq:loglikely}
\end{equation}

\noindent The probability $p_0$ that the structure in the spectrum is due to a background fluctuation is then given by 
\begin{equation}
p_0=1-F(\text{ndf}_{\text{bg}}-\text{ndf}_{\text{bg+s}},\chi_{\text{bg}}^2-\chi_{\text{bg+s}}^2 ) \label{eq:p_0 value}
\end{equation}

\noindent where $F(\Delta_{\text{ndf}}, \Delta \chi^2)$ is the cumulative $\chi^2$ distribution function and $\Delta_{\text{ndf}}$ is the difference in the number of degrees of freedom in both models.

\begin{figure*}
\begin{center}
 \resizebox{1.0\textwidth}{!}
 {\includegraphics[width=8cm]{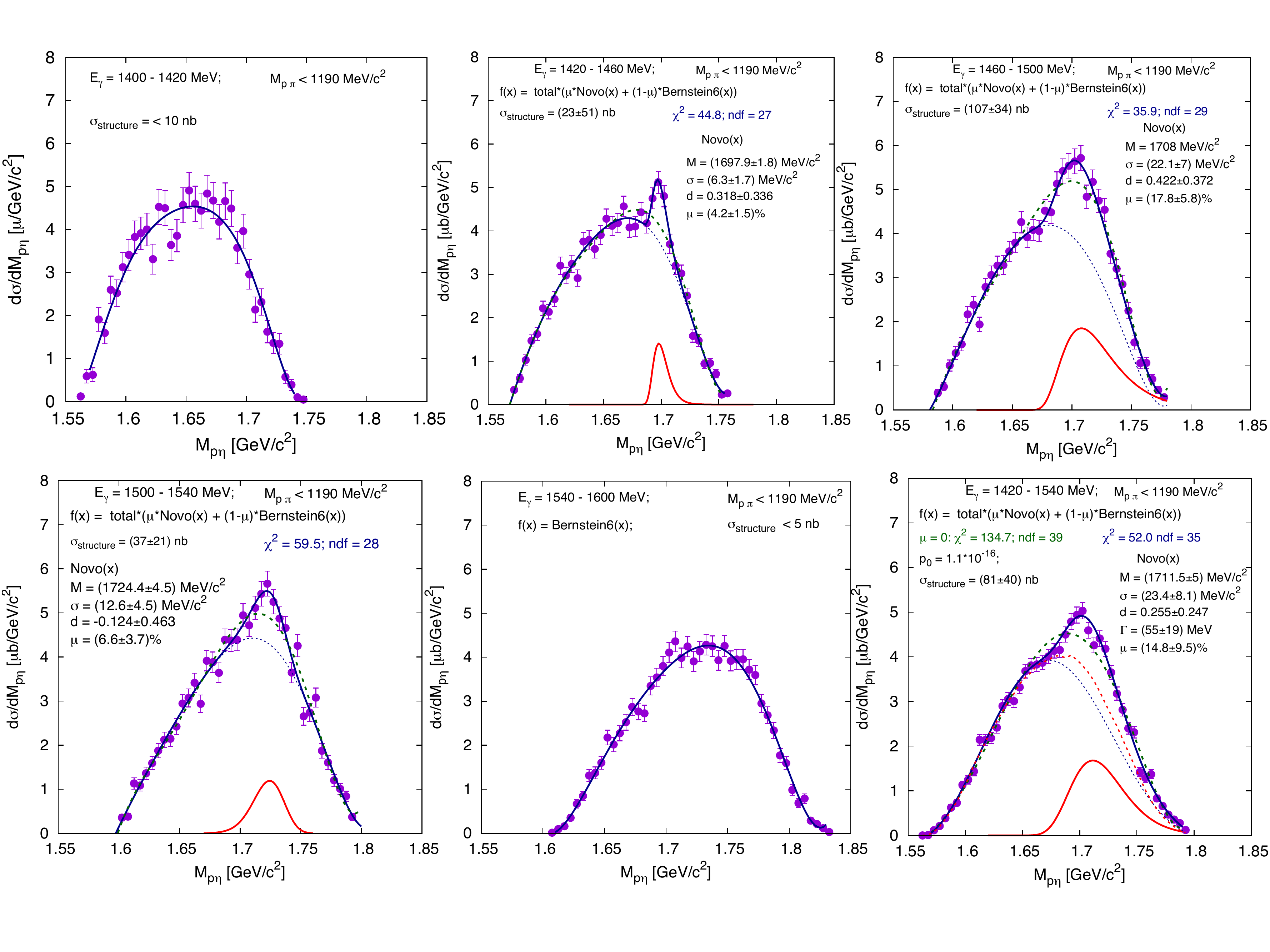}}
 \vspace{-1.2cm}
 \caption{The $M_{p\eta}$ invariant mass distributions for different photon energy bins. The blue solid curves represent fits to the data assuming a Bernstein polynomial of 6th order and a Novosibirsk function for the signal, allowing for asymmetric signal shapes. The dotted blue curves show the fitted background and the red curves the signal, respectively. The green dashed curve corresponds to the fit with the null-hypothesis. For the energy ranges $E_{\gamma}$ = 1400 - 1420 MeV and 1540 - 1600 MeV only an upper limit for the signal can be given. The panel on the bottom right shows the $M_{p\eta}$ spectrum integrated over the relevant energy range $E_{\gamma}$ = 1420 - 1540  MeV. The FWHM of the signal structure is 55 $\pm$ 19 MeV/c$^2$. The red dashed curve represents the $M_{p\eta} $ invariant mass distribution provided by the PWA \cite{Klempt} already shown in Fig.~\ref{fig:comp_PWA_cuts}.}  
\label{fig:structure_energy_dep}
\end{center}
\end{figure*}

\begin{figure*}
\begin{center}
 \resizebox{1.0\textwidth}{!}
 {\includegraphics[width= 10cm]{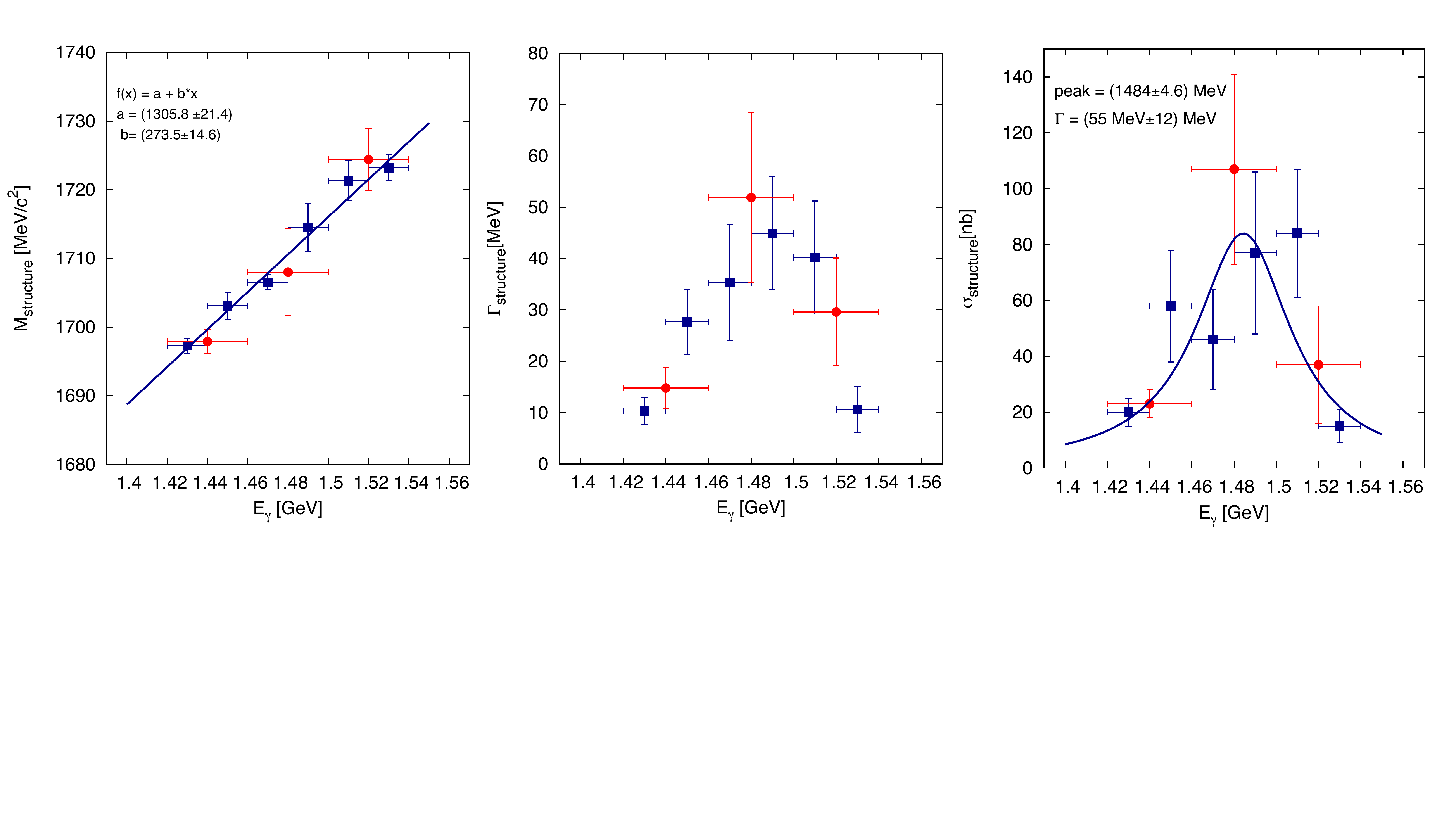}}
 \vspace{-4cm}
 \caption{The peak position (left), the width (middle) and the cross section (right) of the structure as a function of the incident photon energy. The blue (red) points represent the results obtained by fits with a Gaussian (Novosibirsk) signal shape, respectively.  The horizontal error bars represent the bin widths. Blue solid curves are fits to the data. The cross section data have been fitted with a relativistic Breit-Wigner distribution.}
\label{fig:peak_width_yield}
\end{center}
\end{figure*}

\begin{figure*}
\begin{center}
 \resizebox{1.0\textwidth}{!}
{\includegraphics[width=6cm]{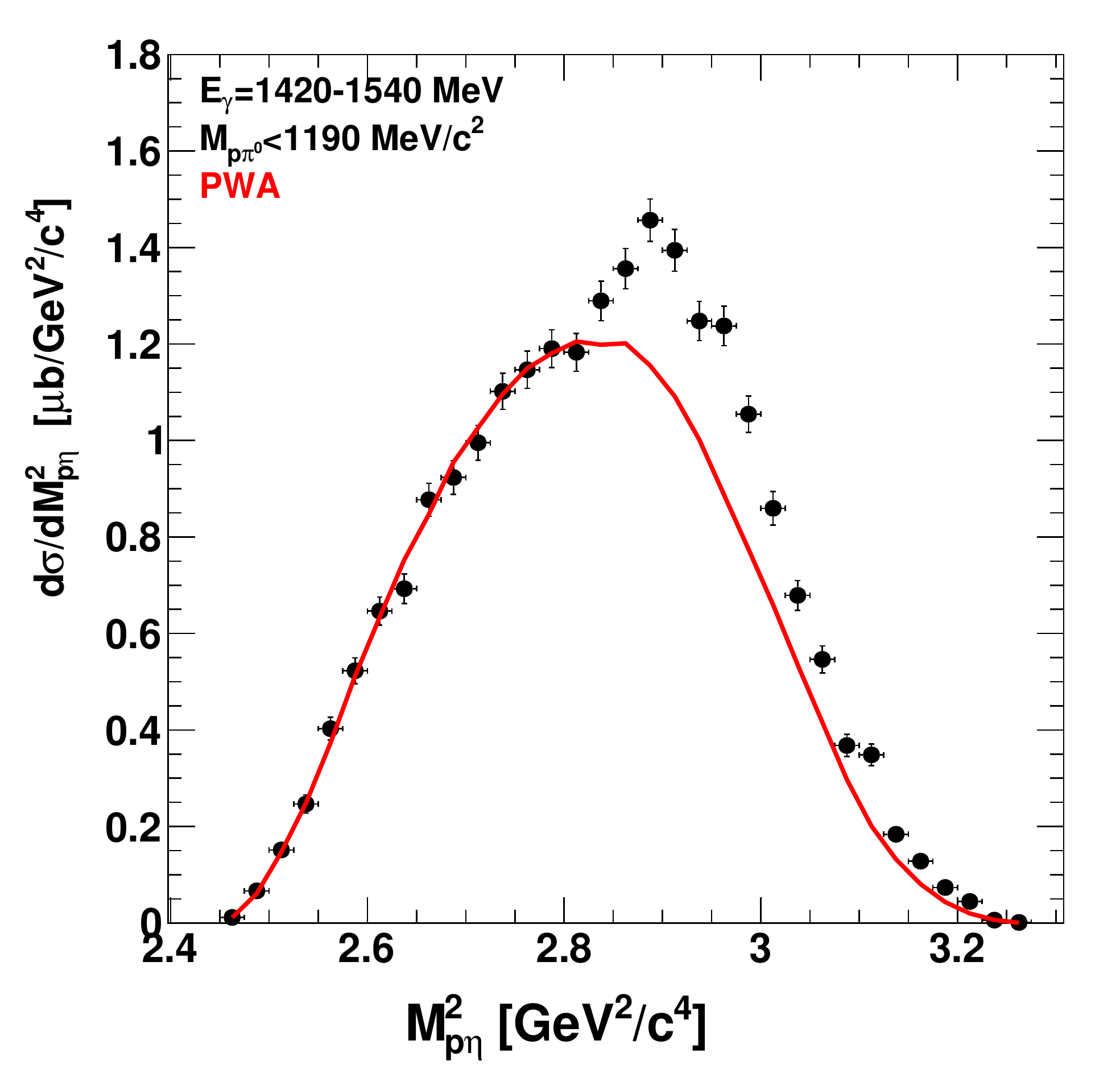}\includegraphics[width=6cm]{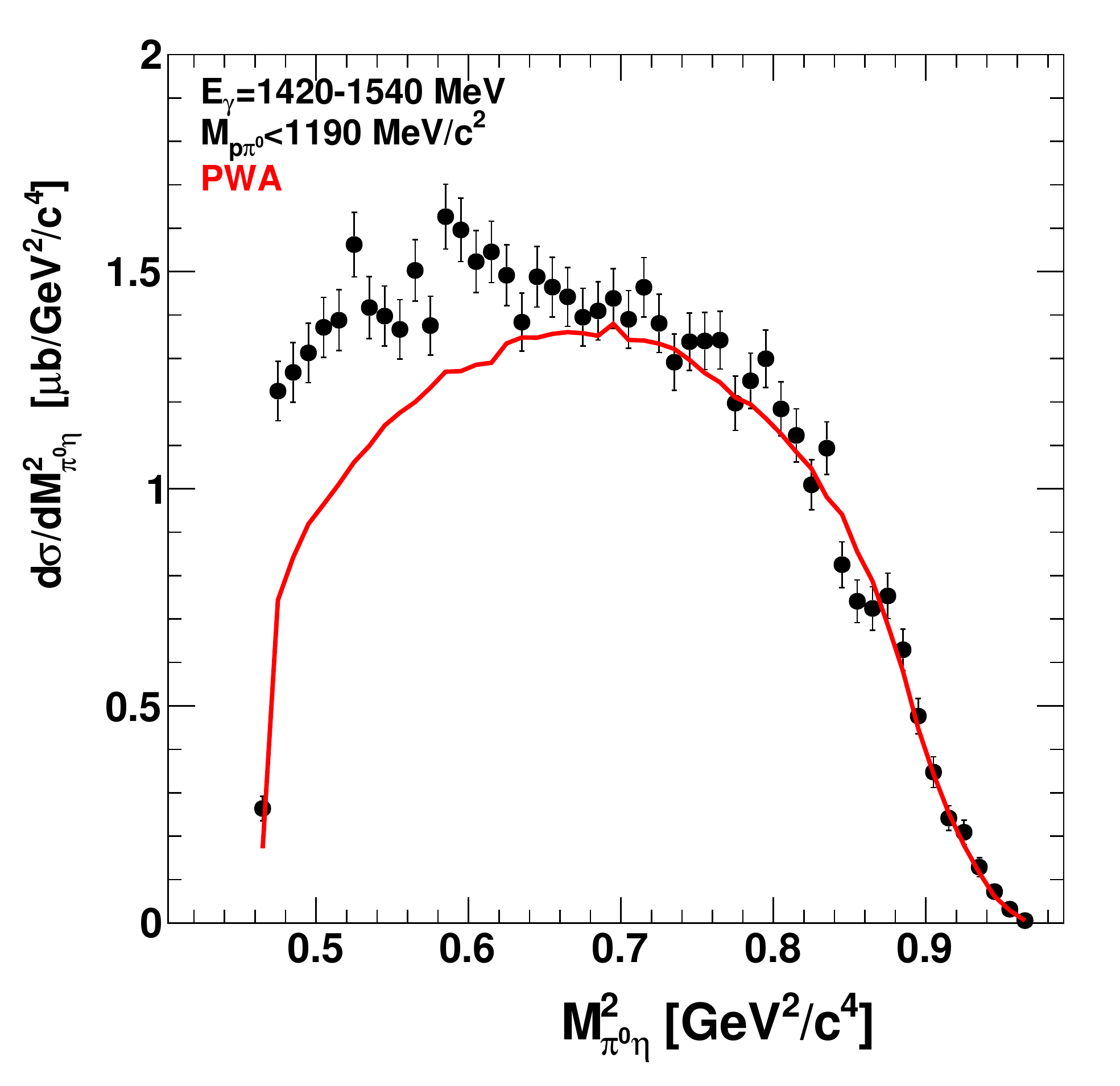}\includegraphics[width=6cm]{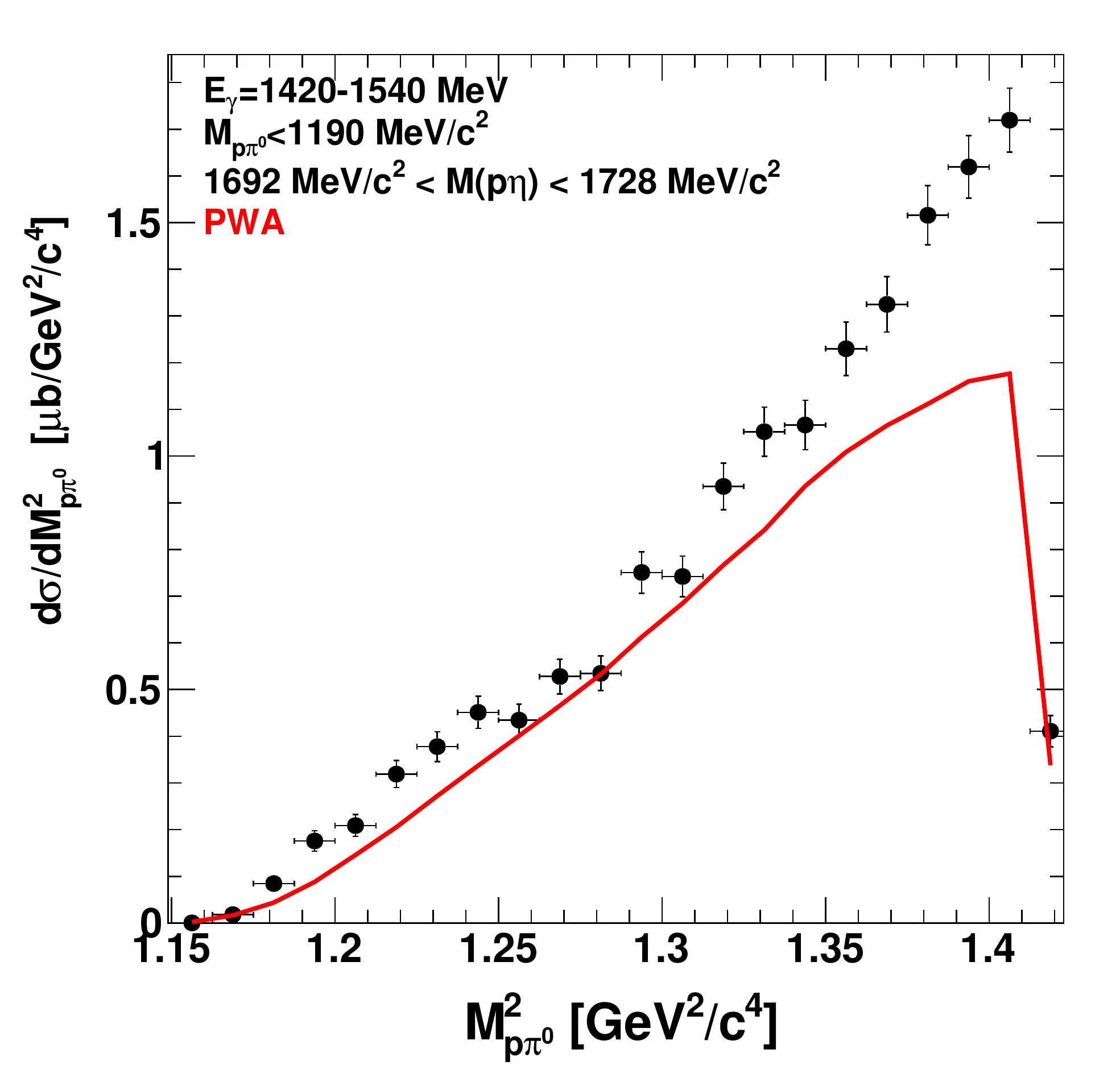}}
 \caption{Excess yield relative to the PWA distributions in the invariant mass distributions $M_{p\eta}^2 $(left), $M_{\pi^0\eta}^2 $ (middle), and $M_{p\pi^0}^2$, the latter with a cut on the structure in the $M_{p\eta}$ invariant mass distribution (1692 MeV/c$^2 \le M_{p\eta} \le $1728 MeV/c$^2$).}
\label{fig:dalitz_proj}
\end{center}
\end{figure*}

\begin{figure*}
\begin{center}
 \resizebox{0.8\textwidth}{!}
{\includegraphics[width = 6cm]{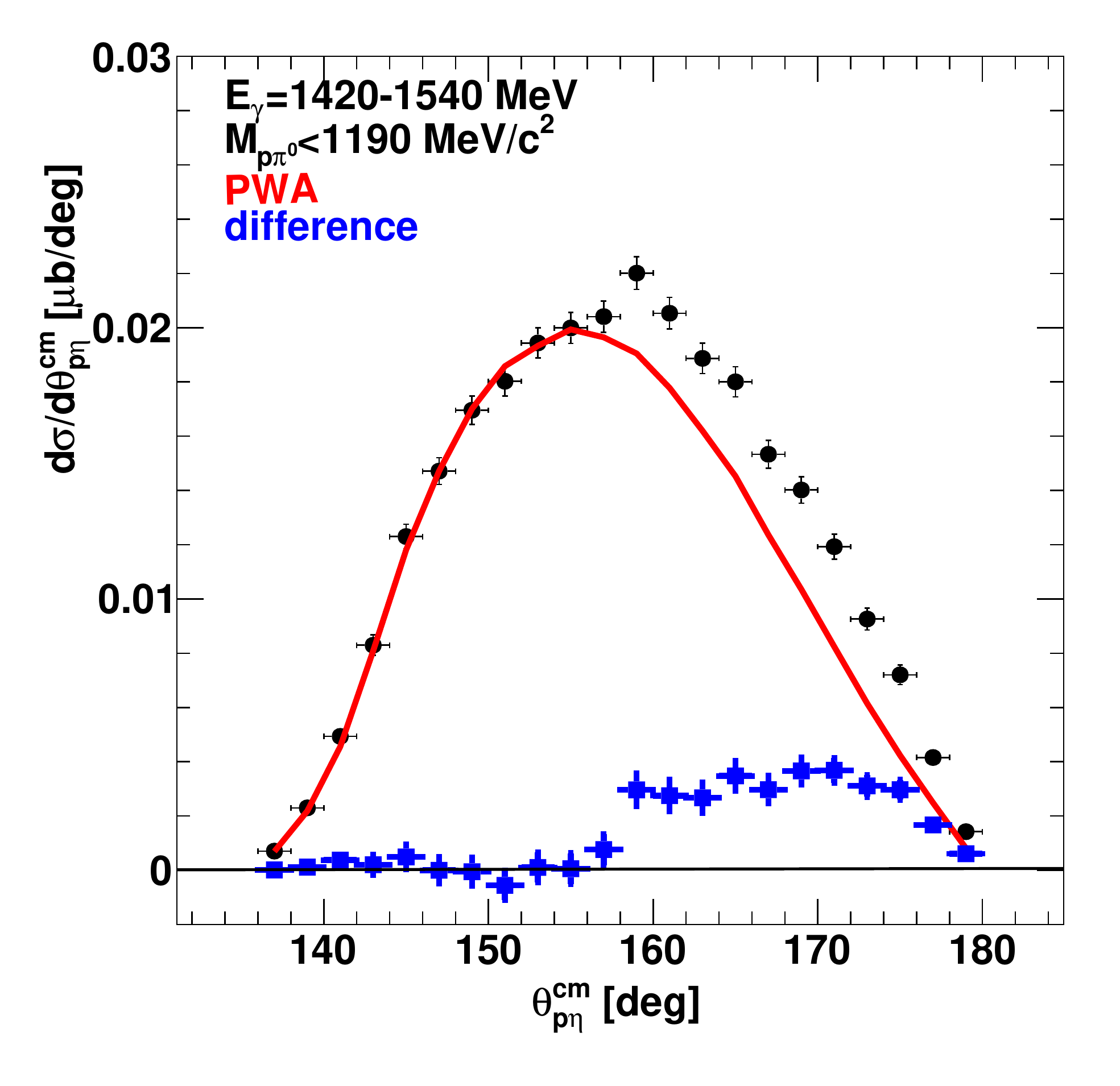}\includegraphics[width = 6cm]{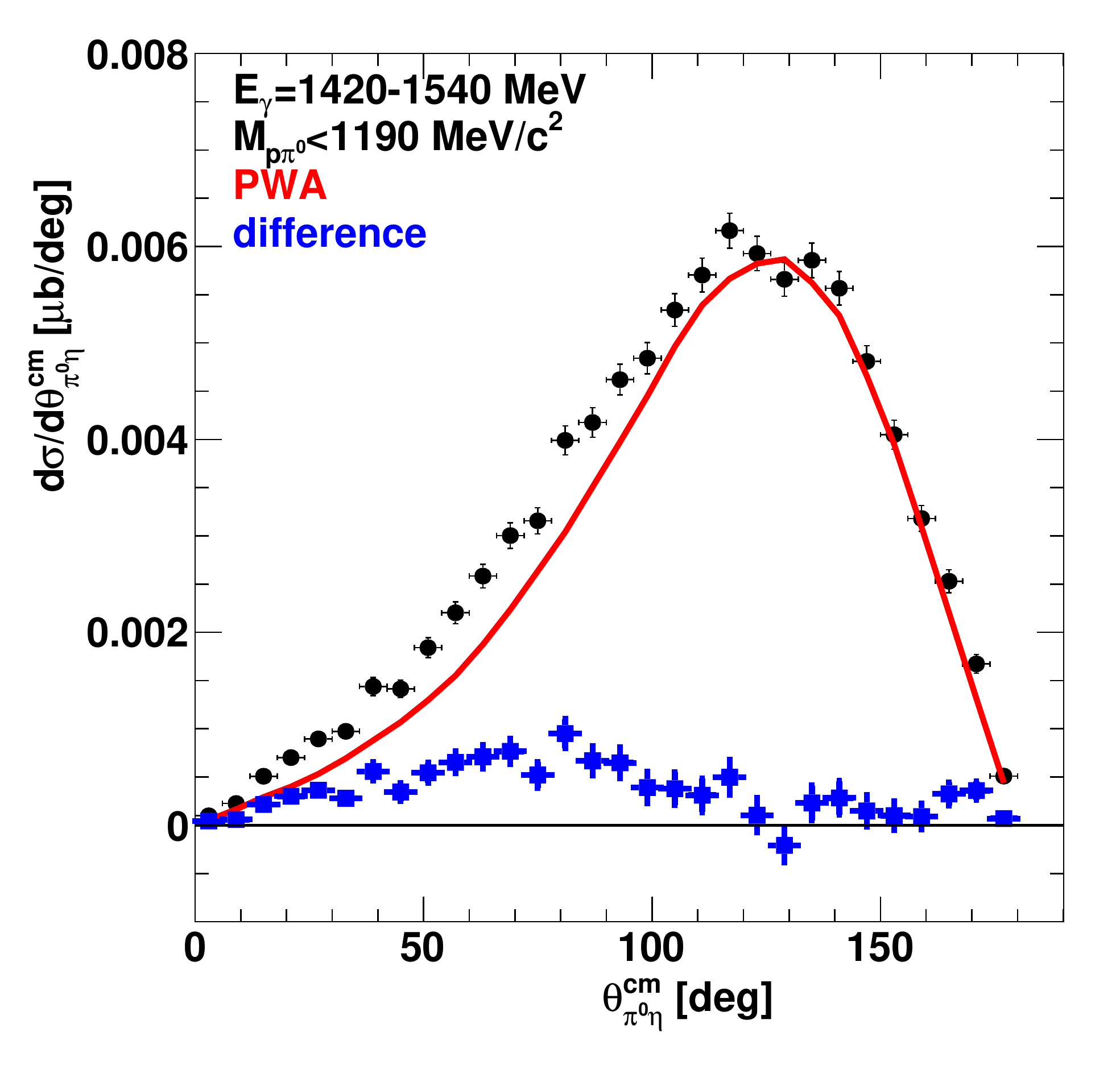}}
 \caption{Excess yield relative to the PWA distributions in the $p-\eta $ (left) and $\pi^0-\eta$ (right) opening angle distributions in the $\gamma p$ centre-of-mass system.}
\label{fig:opening_angle}
\end{center}
\end{figure*}

With $\chi^2_{\text{bg}}$ = 90.0; ndf$_{\text{bg}} = 37$ and $\chi^2_{\text{bg+s}}= 41.9; $ ndf$_{\text{bg+s}}$ = 34, obtained in the fits to the spectrum in Fig.~\ref{fig:structure} left, the probability of the observed structure to arise from a background fluctuation is $p_0 = 2.0\cdot10^{-10} $. The null-hypothesis is thus rejected; the equivalent significance is 
Z = $\Phi^{-1}(1-p_0)$ = 6.5  $\sigma$ (hereby $\Phi^{-1} $ is the inverse of the cumulative distribution of the standard Gaussian function). The statistical significance of the structure has further been studied by determining the local $p_0$ value at a given $M_{p\eta} $ mass, varied in steps of 1 MeV/c$^2$, as shown in Fig.~\ref{fig:structure} (right). The largest local significance is found for $M_{p\eta} = 1700.0 $ MeV/c$^2$ where it reaches $p_0 =  3.9\cdot10^{-12} $ corresponding to 6.8 $\sigma$. The existence of this structure is therefore clearly established.

The structure is not caused by the $M_{p\pi} $ cut as demonstrated in Fig.~\ref{fig:comp_PWA_cuts} (top, right) where the experimental $M_{p\eta} $ distribution is compared with the PWA and phase space simulations. All three distributions are subject to the cut $M_{p\pi} \le 1190 $~MeV/c$^2$. The two simulations which differ in the distribution of events in momentum space do not exhibit a structure as observed in the experimental spectrum. Thus the structure is not a consequence of the $M_{p\pi} $ cut. Varying the $M_{p\pi} $ cut as in the other plots of Fig.~\ref{fig:comp_PWA_cuts}, all $M_{p\eta} $ distributions exhibit an extra strength relative to the PWA distribution in the $M_{p\eta}$ mass range of about 1680 - 1750~MeV/c$^2$, indicating  physics not contained in the current partial wave analysis. The peak position around $M_{p\eta} \approx 1700 $~MeV/c$^2$ is rather stable against varying the $M_{p\pi} $ cut, but the most pronounced structure is found for $M_{p\pi} \le 1190 $~MeV/c$^2$. All plots in Fig.~\ref{fig:comp_PWA_cuts} are shown for the incident photon energy range $E_{\gamma}$ = 1420 - 1540~ MeV which is investigated in the subsequent sections to clarify the nature of the observed structure.

\subsection{Properties of the structure as function of the incident photon energy}
\label{E_dep_structure}

Information on the nature of the observed structure may be obtained by studying the signal profile as a function of excitation energy. Fig.~\ref{fig:structure_energy_dep} shows $M_{p \eta}$ invariant mass distributions for the cut $M_{p \pi^0} \le $1190 MeV/c$^2$ for different bins in the incident photon energy range of $E_{\gamma}$ = 1400 - 1600 MeV. The  spectra have been fitted with a Bernstein polynomial of 6th order and a Novosibirsk function \cite{Aubert} to allow for asymmetric line shapes. Alternatively a relativistic Breit-Wigner distribution and a Gaussian function have been used to describe the signal (not shown). Results obtained for the widths and yields using the different fit functions agree within $\approx 15\%$ which is taken as the systematic fit error (s. table \ref{tab:syst}); the statistical errors are given in the figures. All fits show a shift in mass and increase in width with increasing incident photon energy: the peak of the structure shifts from 1698 MeV/c$^2$ at $E_{\gamma} $= 1420-1460 MeV to 1724 MeV/c$^2$ at $E_{\gamma} $= 1500 -1540 MeV (s. Fig.~\ref{fig:peak_width_yield} (left)). The width increases from FWHM $\approx$ 15 MeV/c$^2$ to FWHM $\approx$  50 MeV/c$^2$ for $E_{\gamma} $= 1460-1500 MeV and then decreases again for higher incident photon energies as shown in Fig.~\ref{fig:peak_width_yield} (middle). The fit of the $M_{p\eta}$ spectrum integrated over the relevant energy range $E_{\gamma}$ = 1420 - 1540  MeV (Fig.~\ref{fig:structure_energy_dep} (bottom, right)) shows an asymmetric line shape with a tail towards larger $M_{p\eta}$ masses. The FWHM of the signal structure is (55 $\pm$ 19) MeV/c$^2$. While the contribution from cascades via the $\Delta(1232) $ resonance shows a rather constant plateau level of $\approx 4.2 $ $\mu$b/GeV throughout the considered photon energy range, indicated by the Bernstein polynomial fit curves, the cross section of the structure, as shown in Fig.~\ref{fig:peak_width_yield}, reaches a maximum of about 100 nb for $E_{\gamma} = (1484 \pm 5 )$ MeV, close to the threshold $E_{\gamma} = 1492 $ MeV for the $\gamma p \rightarrow p a_0$ reaction, and falls off at higher energies.  The red dashed curve in Fig.~\ref{fig:structure_energy_dep} (bottom; right) represents the PWA distribution already shown in Fig. ~\ref{fig:comp_PWA_cuts} (top; right). It is not identical but close to the background curve determined by the fit.

The structure can thus be further characterised by analysing the deviation from the PWA calculation in all three invariant mass spectra: $M_{p\eta}^2$, $M_{\pi^0\eta}^2$,  and $M_{p\pi^0}^2$, as shown in Fig.~\ref{fig:dalitz_proj}. The excess yield in $M_{p\eta}^2$ for 2.85 - 3.1 GeV$^2$/c$^4$ is associated with the one in $M_{\pi^0\eta}^2$ below 0.65 GeV$^2$/c$^4$, while the deviation from the PWA in $M_{p\pi^0}^2$ is concentrated near 1.4 GeV$^2$/c$^4 $, cut off to higher invariant masses by the $M_{p \pi^0} \le $1190 MeV/c$^2$ cut. Also the opening angle distributions in the $\gamma p$ centre-of-mass system, shown in Fig.~\ref{fig:opening_angle}, exhibit deviations from the PWA calculations. The $p-\eta$ opening angles associated with the structure in the $M_{p\eta}$ invariant mass are concentrated in the angular range 160$^{\circ} - 180^{\circ} $ indicating a predominant back-to-back emission while the excess yield in the $\pi^0-\eta$ opening angles is found for $\theta_{\pi^0\eta} \approx 20^{\circ} - 90^{\circ}$ degrees. All these characteristic features of the structure presented in this section will have to be reproduced in attempts to reveal the origin of the observed structure.

\section{Discussion}

\subsection{Decay cascade}

A possible scenario is that the structure is populated in the decay of a nucleon resonance with mass of $\approx$ 1930 MeV/c$^2$. Since the mass of the structure is around 1700 MeV/c$^2$ it cannot be populated by $\eta$ emission from this excitation energy range but only by the $\pi^0 $ decay of the initially populated resonance. Subsequently, the structure would then have to decay into a proton and an $\eta$ meson. The well established N(1710) 1/2$^+$ resonance might be a candidate for the observed structure as it is known to have a strong $\eta$-decay fraction of 10-50$\%$ \cite{PDG}. However, according to \cite {PDG} the width of this resonance is 80 - 200 MeV/c$^2$, much larger than the width of the observed structure. Simulations of such a decay cascade have shown that the observable width of the intermediate state may be narrowed by effects of the phase space limit and the $M_{p\pi^0} \le $ 1190 MeV/c$^2$ cut which suppresses high momentum pions in the first decay step. Only assuming unrealistically small widths of the order of 50 MeV/c$^2$ of the initially populated and the intermediate resonance lead to an observable width as found experimentally. Even then it is not possible to reproduce the rather sharply peaked excitation function of the structure with a width of $\le$ 60 MeV/c$^2$ (s. Fig.~\ref{fig:peak_width_yield}). Thus, a standard decay cascade appears to be an unlikely scenario, although interference effects which might produce narrow structures can of course not be excluded. This possibility can only be further pursued in a new partial wave analysis based on the present data which, however, is out of scope of the current work.
\subsection{State in the exotic baryon anti-decuplet}
In subsection \ref{anti_Kuznetsov}  it has been shown that the structure at $M_{p\eta} $ = 1685 MeV/c$^2$ reported by Kuznetsov et al.\cite{Kuznetsov} and associated with the second member of the exotic anti-decuplet cannot be confirmed. Nevertheless, the mass of the structure observed in the present work is still close to that predicted for that state within the Chiral-Soliton Model \cite{Diakonov}. However, the observed energy dependence of the signal is not consistent with such an interpretation. In particular, the data presented in the subsequent section \ref{singularity} correlate the observation of the structure in the $M_{p\eta}$ invariant mass distribution with the opening of the proton-$a_{0}$ channel. Thus, the interpretation as resulting from a triangular singularity  - discussed below - is much more likely than the highly speculative hypothesis of an anti-decuplet state.

\begin{figure*}
\begin{center}
 \resizebox{0.9\textwidth}{!}
 {\includegraphics[width= 8 cm]{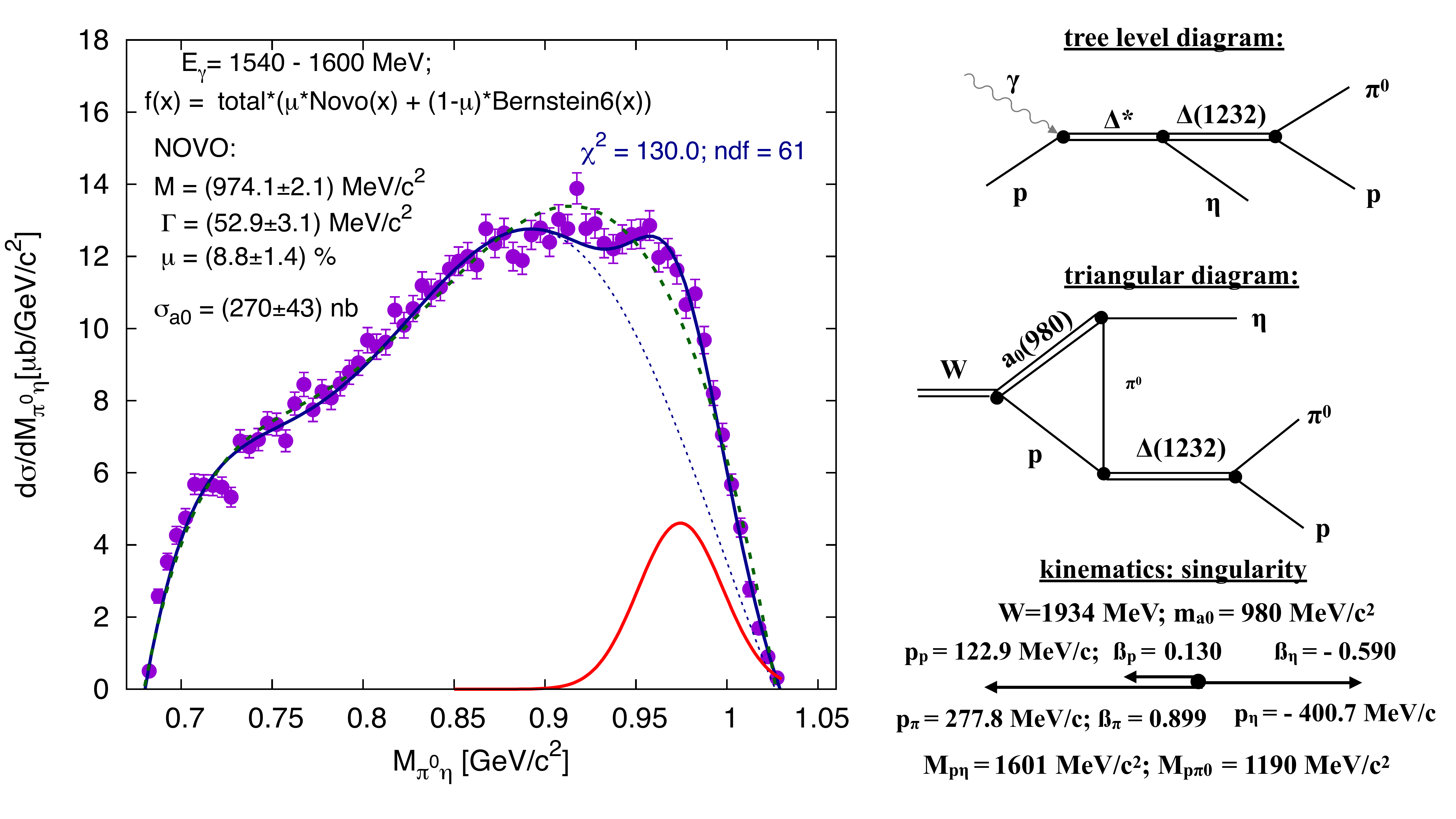}}
 \caption{Left: Differential cross section d$\sigma/dM_{\pi^0\eta}$ for $E_{\gamma}$ = 1540 - 1600 MeV fitted with a Bernstein polynomial of 6th order and a signal described with a Novosibirsk function allowing for asymmetric lines shapes as expected for the $a_{0}$ signal. The line shape of the $a_0 \rightarrow \pi^0 \eta$  signal is complex because of the onset of the $K\bar{K}$ threshold \cite{Flatte,Bugg1,Bugg2}. The $a_0 \rightarrow \pi^0 \eta$ signal line shape is narrowed by the $a_0$ coupling to  $K\bar{K}$ (Flatte shape) but smeared out  by the detector resolution \cite{Bugg2}, justifying the choice of a Novosibirsk signal profile. The dotted blue curve shows the fitted background and  the red curve the signal, respectively. The green dashed curve corresponds to the fit with the null-hypothesis. (Right, top): tree level diagram describing the dominant  $\gamma p \rightarrow \Delta^* \rightarrow \eta \Delta(1232) \rightarrow \eta p \pi^0$ reaction channel; (right, middle): three-point loop diagram with $a_0, \pi^0 $ and $p$ as intermediate particles. The proton and the $\pi^0$ from the $a_0 \rightarrow \pi^0 \eta$ decay fuse to form the $\Delta(1232)$ resonance subsequently decaying into $p$ and $\pi^0$. (Right, bottom) The collinear kinematics in the $\gamma$p centre-of-mass system associated with the triangular singularity at W = 1934 MeV and $m_{a0}$ = 980 MeV/c$^2$. }
 \label{fig:a_0}
\end{center}
\end{figure*}

\begin{figure*}
\begin{center}
 \resizebox{1.0\textwidth}{!}
{\includegraphics[width=10 cm]{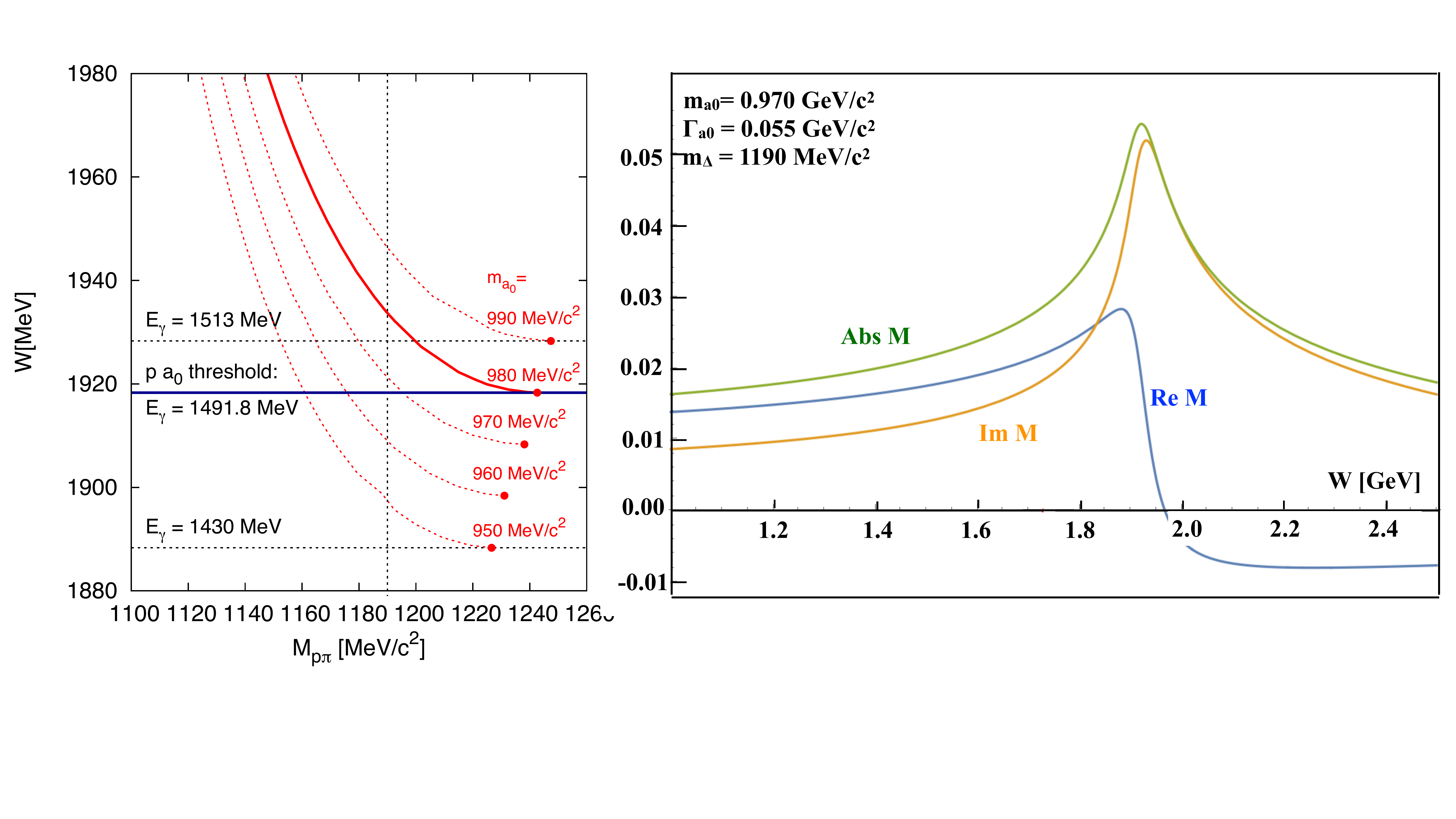}}
\vspace{-2cm}
 \caption{Left: Correlation between the excitation energy W and the invariant mass $M_{p\pi^0}$. The red curves represent the calculated location of singularities as function of excitation energy W and $M_{p\pi^0}$ for different values of the $a_{0}$ mass close to its pole. The cut $M_{p\pi^0} \le$ 1190 MeV/c$^2$ is indicated by a dashed vertical line. The red stars mark the singularities used in the calculations shown in Figs. \ref{fig:Dalitz-rescattering} - \ref{fig:open_ang_calc}. The horizontal dashed lines correspond to the thresholds for $a_0$ masses of 950 MeV/c$^2$ and 990 MeV/c$^2$, respectively, and mark the incident photon energy range of interest. Right: The real part, imaginary part and total triangular amplitude, calculated for $m_{a_0} = 970$ MeV/c$^2$; $\Gamma_{a_0} = 55 $ MeV/c$^2$ and $M_{\Delta} $= 1190 MeV/c$^2$.}
 \label{fig:singularities}
\end{center}
\end{figure*}

\subsection{Triangular singularity}
\label{singularity}
\begin{figure}
\begin{center}
 \resizebox{0.5\textwidth}{!}
{\includegraphics[width=6cm]{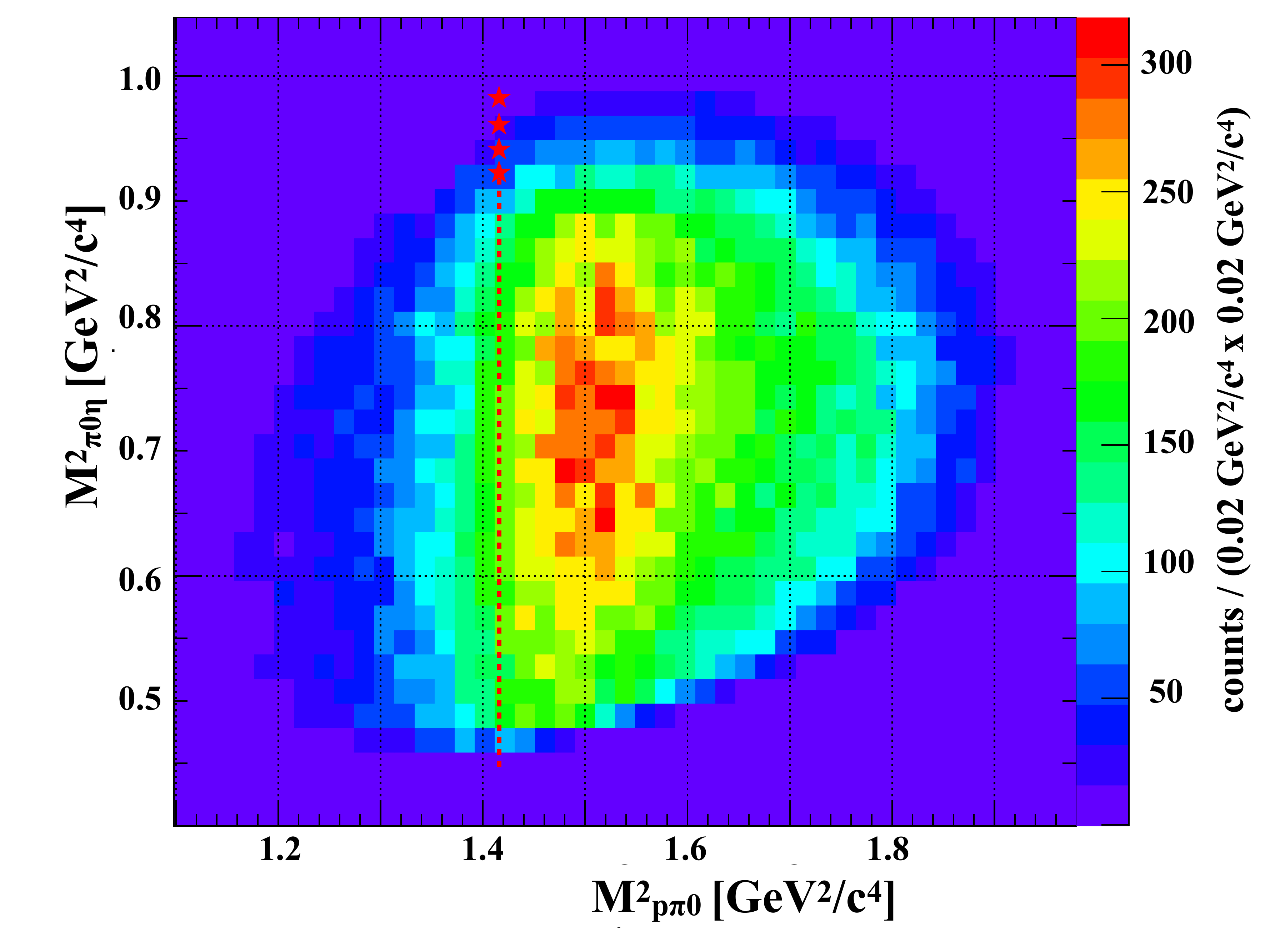}}
 \caption{Dalitz-plot $M_{\pi^0\eta}^2$ versus $M_{p\pi^0}^2$ for $E_{\gamma}$ = 1420 - 1540 MeV. The location of the singularities used in the calculation are indicated by the red stars. Elastic $\pi^0$-p scattering displaces the singularity events along the dotted red line, leading to low $M_{\pi^0\eta}$ masses and high $M_{p\eta}$ masses while $M_{p\pi^0}$ stays unchanged. Because of their small intensity the singularity events cannot be directly seen in the two-dimensional display but show up for $M_{p\pi^0} \le $ 1190 MeV/c$^2$ in the projections (Fig.~\ref{fig:dalitz_proj}) as deviations from the PWA predictions.}  
 \label{fig:Dalitz-rescattering}
\end{center}
\end{figure}

\begin{figure*}
\begin{center}
 \resizebox{1.0\textwidth}{!}
 {\includegraphics[width=8cm]{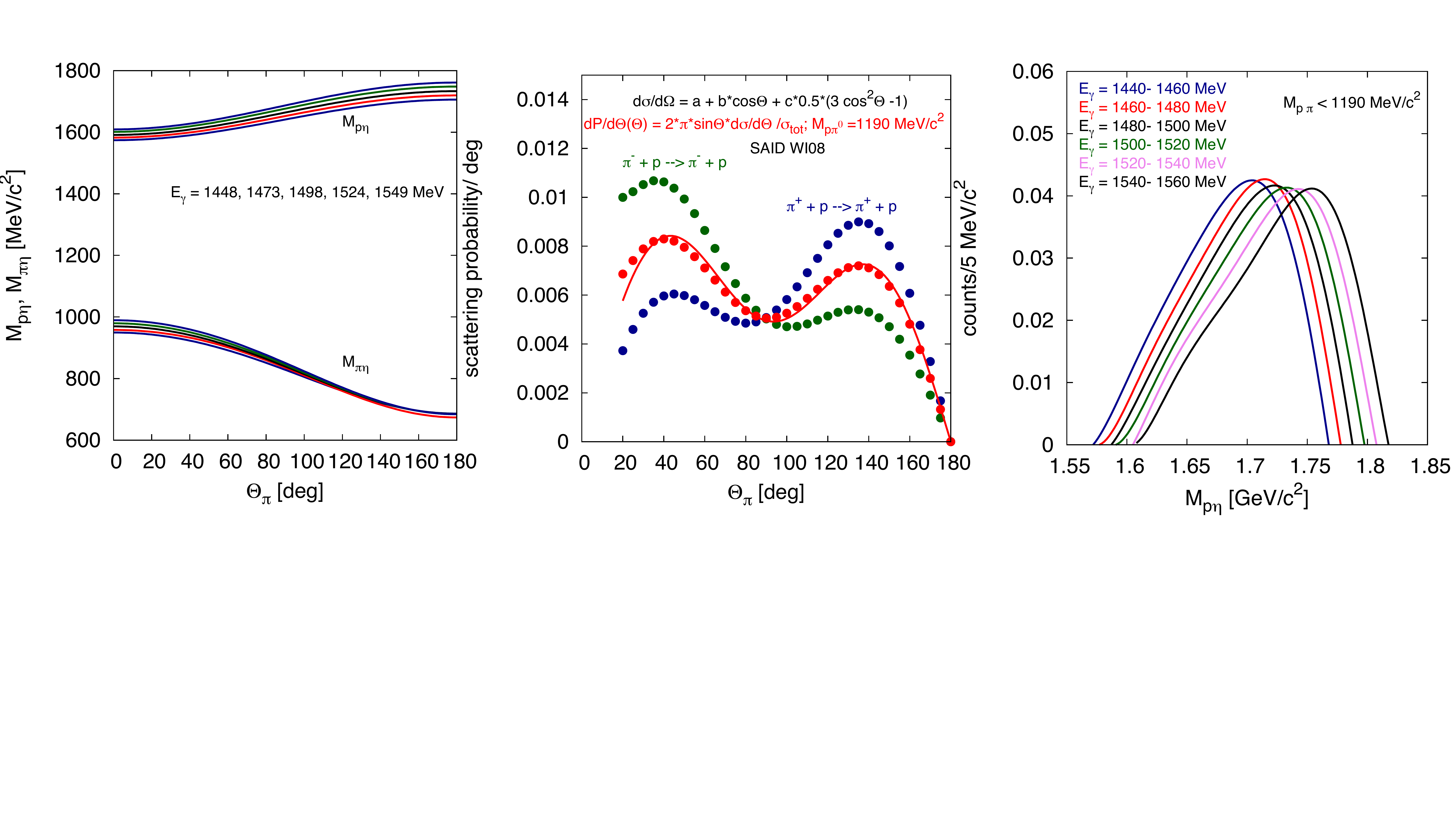}}
 \vspace{-4cm}
  \caption{Left: Invariant masses $M_{p\eta}$ and $M_{\pi^0\eta}$ as function of the $\pi^0$ scattering angle for incident photon energies indicated in the figure. Middle: scattering probability for  $\pi^- p \rightarrow \pi^- p$ (green points) and $\pi^+ p \rightarrow \pi^+ p $ (blue points) at $M_{p\pi}=$ 1190 MeV/c$^2$ in steps of 5$^{\circ}$, taken from \cite{Brack1,Brack2,Ritchie,SAID}. The $\pi^0 p \rightarrow \pi^0 p $ scattering probability in the $\pi^0 - p$ system (red points and fit curve) is approximated by the average of the charged pion scattering probability distributions. Right: $M_{p\eta}$ phase space distributions for $M_{p\pi^0} \le $ 1190 MeV/c$^2$ and the $E_{\gamma}$ ranges given in the figure.}
\label{fig:scattering}
\end{center}
\end{figure*}

Several narrow structures observed in hadron physics have recently been attributed to triangular singularities which are threshold phenomena with special kinematics (for a review see \cite{Guo}). A prominent example is the narrow $a_1$(1420) peak reported by the COMPASS collaboration \cite{COMPASS} which was later shown \cite{Ketzer,Bayar} to arise from a three-point loop in the $a_1^- \rightarrow \pi^- f_0$ decay where the intermediate particles $K^{*0}, K^+ $and $K^- $ are almost on their mass shell, causing a resonance-like effect. As pointed out in section \ref{E_dep_structure}, it is a remarkable feature of the present data that the intensity of the observed structure peaks at $E_{\gamma}$ = 1485 MeV (W = 1918 MeV) near the $a_0$ production threshold. Furthermore, the $M_{\pi^0\eta}$ invariant mass spectrum for the incident photon energy range $E_{\gamma} $ = 1540 - 1600 MeV does indeed exhibit an $a_0 \rightarrow \pi^0 \eta$ signal at $M_{\pi^0\eta} \approx $ 974 MeV/c$^2$ with a width of $\Gamma \approx$ 50 MeV/c$^2$, as shown in Fig.~\ref{fig:a_0}, in good agreement with the PDG values of 980 $\pm$ 20 MeV/c$^2$ and 50 - 100 MeV/c$^2$, respectively \cite{PDG}. The fitted peak position is slightly below the nominal $a_0$ mass, but for incident energies close to the production threshold the low mass side of the $a_0$ mass distribution is preferentially populated and the $a_0 \rightarrow \pi^0 \eta$ branch for higher $a_0$ masses is attenuated due to the opening of the $K \bar{K}$ threshold at $E_{\gamma} $ = 1515 MeV (W= 1930 MeV) \cite {Flatte,Bugg1,Bugg2}. The cross section of (270 $\pm$ 43) nb is in good agreement with $\sigma_{a_0}$ = 250 nb obtained in the PWA  \cite{Gutz} for this incident photon energy range. At higher incident photon energies the $a_0$ signal becomes much more pronounced \cite{Gutz,Celentano}.
\begin{figure*}
\begin{center}
 \resizebox{0.8\textwidth}{!}
 {\includegraphics[width= 10 cm]{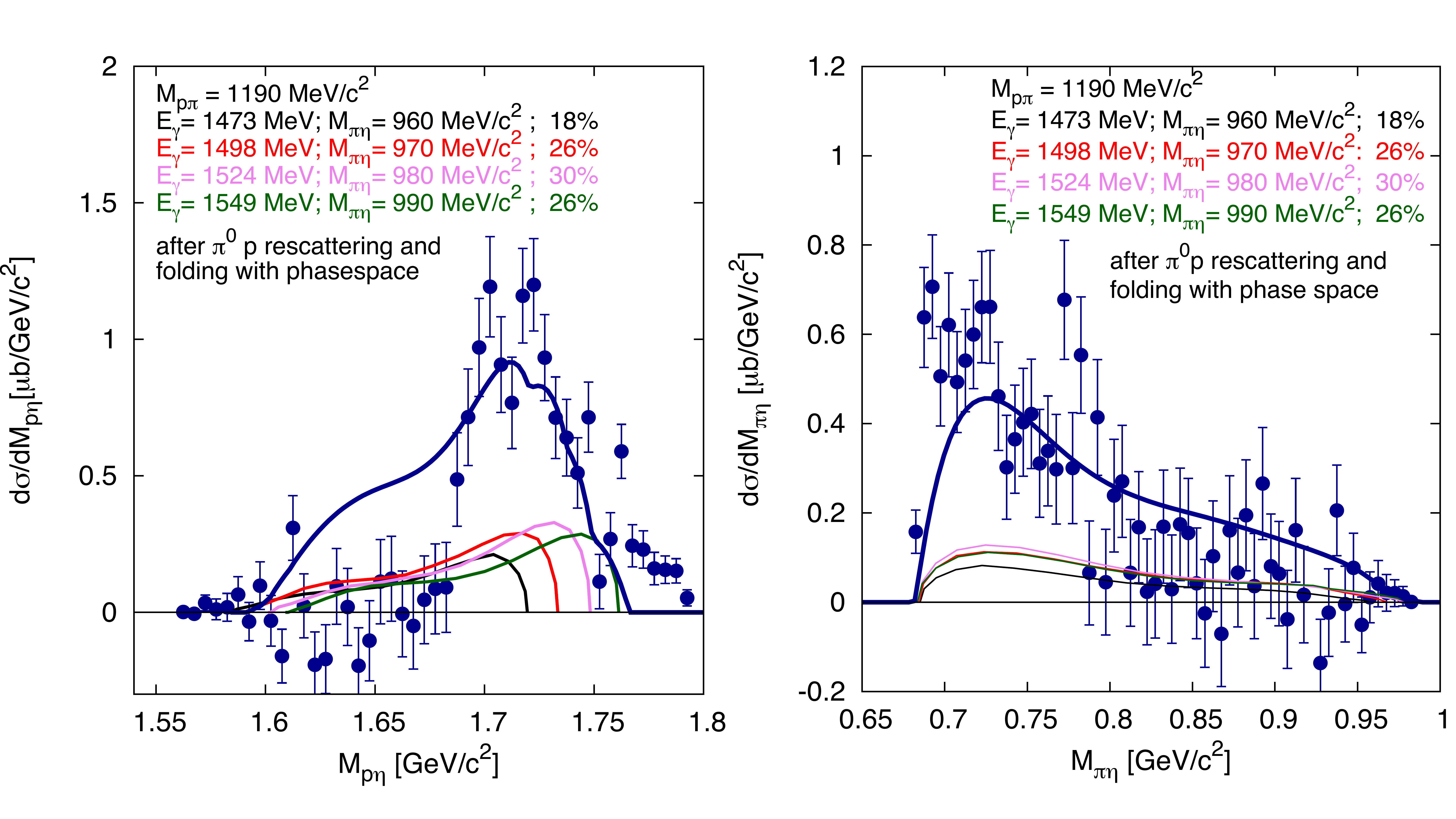}}
 \caption{Calculated $M_{p\eta} $ (left) and $M_{\pi^0\eta} $ (right) invariant mass distributions after $\pi^0$-$p$ re-scattering for different incident photon energies in comparison to the experimental signal, taken as difference to the PWA calculations (s. Fig.~\ref{fig:dalitz_proj}). The thick blue curve, fitted to the data, is the sum of contributions from 4 selected singularity points with relative weight given by the $a_0$ line shape.}
 \label{fig:M_p-eta_calc}
\end{center}
\end{figure*}

\begin{figure*}
\begin{center}
 \resizebox{0.8\textwidth}{!}
 {\includegraphics[width= 10 cm]{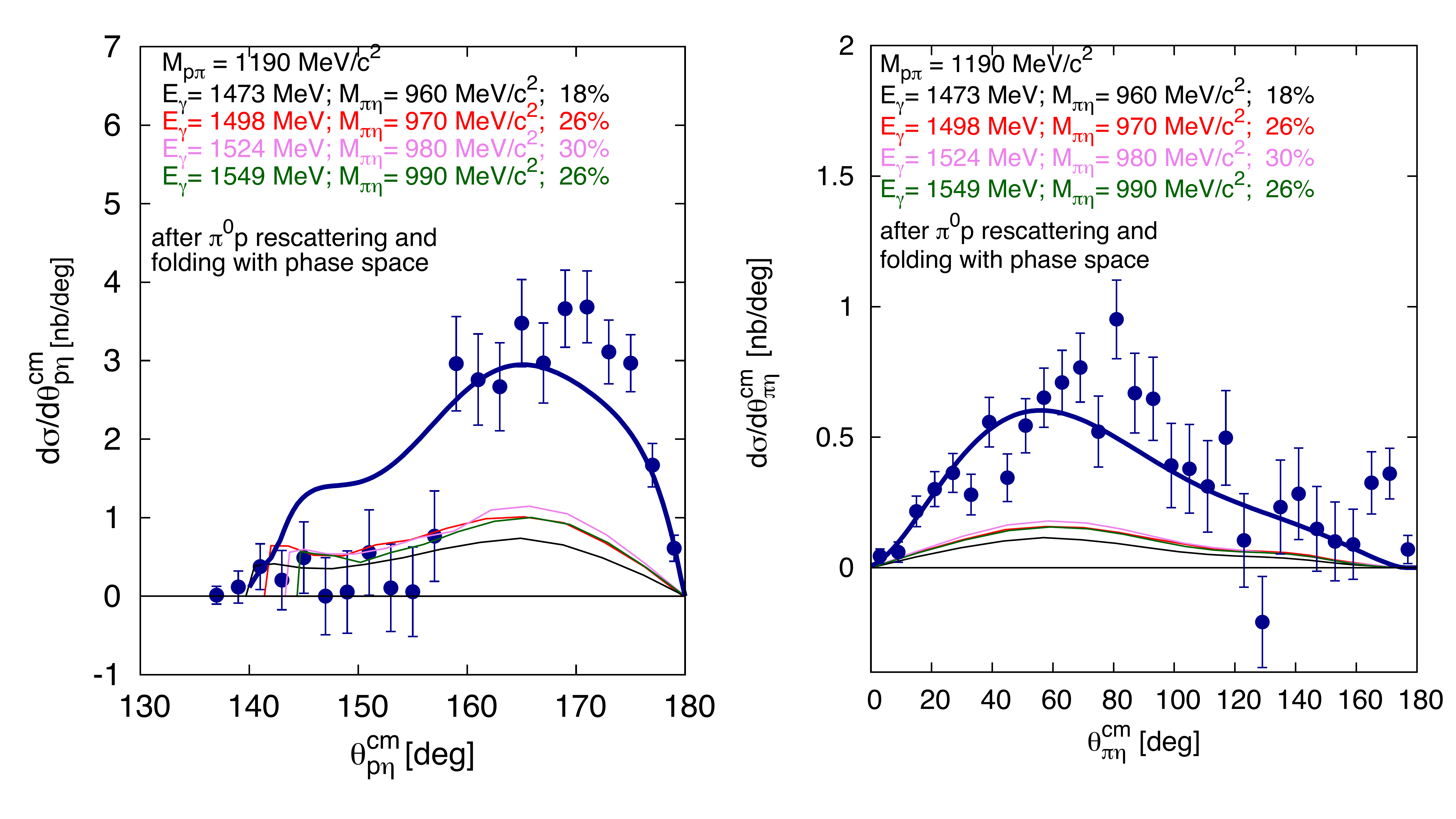}}
 \caption{Calculated opening angle distributions $\theta_{p\eta}$ (left) and $\theta_{\pi^0\eta} $ (right) in the $\gamma$p centre-of-mass system after $\pi^0$-$p$ re-scattering for different incident photon energies in comparison to the experimental signal, taken as difference to the PWA calculations (s. Fig.~\ref{fig:opening_angle}). The thick blue curve, fitted to the data, is the sum of contributions from 4 selected singularity points with relative weight given by the $a_0$ line shape.}
 \label{fig:open_ang_calc}
\end{center}
\end{figure*}

The observation of the $a_0$ signal suggests to investigate the possibility of a triangular singularity in the $a_0 \rightarrow \pi^0 \eta$ decay as origin of the observed structure. Thus, in addition to the tree level diagram (s. Fig.~\ref{fig:a_0} (right, top)) describing the dominant  $\gamma p \rightarrow \Delta^* \rightarrow \eta \Delta(1232) \rightarrow \eta p \pi^0$ reaction channel a three-point loop diagram depicted in Fig.~\ref{fig:a_0} (right, middle)) has to be considered. As outlined in \cite{Guo,Ketzer,Bayar,Landau,Schmid,Aceti,Xie_Guo,Debastiani} this loop diagram develops a singularity whenever the energy and momentum balance of the initial and final state particles matches the energy balance of the intermediate particles within the loop with all particles almost on-shell. This condition can only be fulfilled for specific kinematic conditions, i.e. for certain pairs of W and $M_{p\pi^0}$ values, as illustrated in Fig.~\ref{fig:singularities} (left). For given $a_0$ masses near the pole of 980 MeV/c$^2$ the energy and momentum balance matching occurs along the dashed red curves indicating the positions of the singularities. Singularities are found at W $\approx$ 1900 - 1960 MeV for $m_{a^0}$ = 960 - 990 MeV/c$^2$ and $m_{\Delta} $ = 1170 - 1232 MeV/c$^2$, respectively. When comparing to the data, however, one cannot get closer to the pole of the $\Delta(1232)$ resonance than $M_{p\pi^0}$ = 1190 MeV/c$^2$ because of the cut applied on the data. Thus for the calculations described in the following, four points at $M_{p\pi^0} $ = 1190 MeV/c$^2$, marked by red stars, have been selected. Calculations of the triangular amplitude as in \cite{Ketzer} - e.g.,  for the point ($m_{a_0}$ = 970 MeV/c$^2$; $m_{\Delta} $ = 1190 MeV/c$^2$) -  lead to a resonance-like enhancement in the $\gamma p \rightarrow p \pi^0 \eta$ cross section shown in Fig.\ref{fig:singularities} (right) to be compared to the experimental data in Fig.~\ref{fig:peak_width_yield} (right).

In the special kinematic condition of the singularity all particles have to be collinear. Following the prescription by \cite{Bayar} the kinematic configuration in the centre-of-mass system shown in Fig.~\ref{fig:a_0} (right, bottom) has been calculated for the case W = 1934 MeV and  $m_{a_0} $= 980 MeV/c$^2$: the $\pi^0$ meson is faster than the proton; it will catch up with the proton and re-scatter, forming a slightly off-shell $\Delta(1232)$ resonance which subsequently decays again into a $\pi^0$ and a proton. This re-scattering process is essential for the interpretation of the observed structure in the invariant mass distribution at $M_{ p\eta} \approx$ 1700 MeV/c$^2$. The four selected singularity points of Fig.~\ref{fig:singularities} (left) are again shown as red stars in the Dalitz-plot $M^2_{\pi^0\eta}$ vs. $M^2_{p\pi^0} $ in Fig.~\ref{fig:Dalitz-rescattering}. These points correspond to $M_{p\eta} \approx$ 1580 - 1610 MeV/c$^2$.  However, in the elastic $\pi^0$-p scattering the singularity events are re-distributed along the dotted red straight line in Fig.~\ref{fig:Dalitz-rescattering}: the $\pi^0$ is slowed down and the proton speeds up, thereby increasing the $M_{p\eta}$ mass to values around 1700 MeV/c$^2$ (the mass range of the structure) and reducing the $M_{\pi^0\eta}$ invariant mass almost to the kinematical limit $m_{\pi^0}$ + $m_{\eta} $ =  683 MeV/c$^2$, as depicted in Fig.~\ref{fig:scattering} (left),  while $M_{p\pi^0}$ stays constant. 

The elastic $\pi^0 p$ scattering cannot be measured but may be estimated from the known differential cross sections for the $\pi^- p \rightarrow \pi^- p $ and $\pi^+ p \rightarrow \pi^+ p $ reactions \cite{Brack1,Brack2,Ritchie,SAID}. At forward angles and for invariant masses below the $\Delta(1232) $ pole mass the charged pion scattering is affected by Coulomb-nuclear interferences which are constructive and destructive, respectively. The differential cross section for $\pi^0 p$ scattering may thus be approximated by taking the average of the two normalized charged pion angular distributions since the Coulomb-nuclear interferences may cancel (s.  Fig.~\ref{fig:scattering} (middle)). 
The $\pi^0 -p$  re-scattering probability as function of the scattering angle $\theta_{\pi^0}$ in the $\pi^0 - p$ system can then be estimated by
\begin{equation}
dP/d{\theta_{\pi^0}} = 2 \pi \sin{\theta_{\pi^0}} \cdot d{\sigma}/d{\Omega}/{\sigma_{tot}} 
\end{equation}
The $M_{p\eta}$ and $M_{\pi^0\eta} $ invariant mass distributions associated with the triangular singularity and the corresponding opening angles in the center-of-mass (cm) system can then be calculated by folding the $\pi^0 - p$ scattering probability with the phase space distribution for $M_{p\pi^0} \le $ 1190 MeV/c$^2$ (s. Fig.~\ref{fig:scattering} (right)). While the scattering probability is almost symmetric around 90$^{\circ}$, the phase space distributions favour high $M_{p\eta}$ masses and thus small $M_{\pi^0\eta}$ masses which are both associated with large scattering angles (s. Fig.~\ref{fig:scattering} (left)).
 
The $M_{p\eta}$ invariant mass distributions calculated for the four selected singularity points are shown in Fig.~\ref{fig:M_p-eta_calc} (left) in comparison to the experimental signal taken as difference of the data to the PWA calculation. The solid blue curve represents the sum of these contributions with a weight given by the $a_0$ line shape, determined in Fig.~\ref{fig:a_0}. Since the calculations only provide the shape of the distributions their sum is fitted to the experimental cross section. As a result of the re-scattering, invariant $M_{p\eta}$ mass distributions are obtained with widths of about 50 MeV/c$^2$ peaking at $M_{p\eta} \approx$ 1700 - 1750 MeV/c$^2$, shifting in mass with increasing incident photon energy, as observed experimentally. After $\pi^0 - p$ scattering $M_{\pi^0\eta}$ masses (Fig.~\ref{fig:M_p-eta_calc} (right)) concentrate close to the kinematical limit $m_{\eta}$+$m_{\pi^0}$ = 682.9 MeV/c$^2$. The calculated $\theta_{p\eta}$ and $\theta_{\pi^0\eta}$ opening angle distributions in the centre-of-mass system are shown in Fig.~\ref{fig:open_ang_calc} in comparison to the distributions of the blue data points in Fig.~\ref{fig:opening_angle}. The $\theta_{p\eta}$ opening angles are confined to the angular range 150$^{\circ}$-$180^{\circ}$ while the $\theta_{\pi^0\eta}$ opening angles exhibit a broad distribution peaking around 60$^{\circ}$ in good agreement with the experimental data. Again the contributions calculated for the different selected singularity points are shown separately while the thick blue curves, fitted to the data, represent the sum of these contributions. 

There is overall a qualitative agreement between data and calculations. Starting from the mathematically proven triangular singularity the characteristic features of the observed structure displayed in Fig.~\ref{fig:structure_energy_dep} to Fig.~\ref{fig:opening_angle} are thus all reproduced: the shift of the peak position with increasing incident photon energy, the limited width $\le$ 50 MeV, the enhancement in yield near $E_{\gamma} \approx $ 1490 MeV (W~\ $\approx$ 1920 MeV), the invariant mass distributions of Fig.~\ref{fig:dalitz_proj}, showing deviations from the current PWA solution at $M_{p\eta} \approx$ 1700 - 1750 MeV/c$^2$, at  low $M_{\pi^0\eta} $ masses and at $M_{p\pi}$ near 1190 MeV/c$^2$, and finally the $\theta_{p\eta}$ and $\theta_{\pi^0\eta}$ opening angle distributions. A quantitative agreement in all details cannot be expected, however, since the present calculation has only been performed for selected points in Fig.~\ref{fig:singularities} (left) to demonstrate that the structure in the $M_{p\eta}$ invariant mass distribution may arise from the singularity. A full calculation would request a continuous coverage of this plane for $M_{p\pi^0} \le $ 1190 MeV/c$^2$ - and more importantly - the interference of the tree- and triangular-diagrams has to be considered. This is out of scope of the present work but will have to be taken into account in a forthcoming partial wave analysis. The current analysis, however, demonstrates the role of triangular loops and the importance of re-scattering effects in the interpretation of structures in the excitation energy spectrum of the nucleon. The present data may also be of interest in view of current discussions of the two-pole structure of nucleon resonances \cite{Khemchandani,Meissner}.

\section{Summary and conclusions}
The $\gamma p \rightarrow p \pi^0 \eta$ reaction has been analysed for incident photon energies of 1400 - 1600 MeV. Under identical conditions as used by Kuznetsov et al. \cite{Kuznetsov}, the analysis of the present data does not confirm their observation of a structure in the $M_{p\eta}$ invariant mass distribution at 1678 MeV/c$^2$. Instead, a structure in the $M_{p\eta}$ invariant mass distribution near 1700 MeV/c$^2$ with a width of $\Gamma \approx 50 $ MeV/c$^2$ is observed for $E_{\gamma}$ = 1420 - 1540 MeV and the cut $M_{p\pi^0} \le 1190 $ MeV/c$^2$. With increasing incident photon energy this structure is found to shift in mass, the width varies but stays below 60 MeV/c$^2$ and the intensity peaks with a cross section of $\approx$ 100 nb around $E_{\gamma} \approx$ 1490 MeV (W $\approx$ 1920 MeV), close to the threshold for the $\gamma p \rightarrow p a_0$ reaction. A scenario, assuming a so far unobserved decay chain via an intermediate resonance near 1710 MeV/c$^2$ does not reproduce the excitation function and the narrow width of the structure, unless interference effects play an important role. In view of the data presented in this work an interpretation as an anti-decuplet state predicted within the Chiral Soliton Model appears highly unlikely. The characteristic features of the observed structure are consistent with a triangular singularity at W~$\approx$~1900 - 1920 MeV, arising from a three-point loop in the $\gamma p \rightarrow p a_0 \rightarrow p\pi^0 \eta$ reaction with the $a_0, \pi^0 $ and proton as intermediate particles. Hereby, the $\pi^0$-$p$ re-scattering is essential for explaining the mass of the observed structure. It will be interesting to see whether the conclusions of the present work can be confirmed in an updated partial wave analysis which includes the new data and all possible interference effects. 

\section{Acknowledgements}
We thank the scientific and technical staff at ELSA and the collaborating institutions for their important contribution to the success of the experiment. Discussions with H.~Clement, A.~Fix, E.~Klempt, W.~K\"uhn, E.~Oset, \\I.~I.~Strakovsky and R.~L.~Workman on various aspects of the analysis and on the  possible interpretation of the experimental results are highly acknowledged. This work was supported financially by the {\it Deutsche Forschungsgemeinschaft} within the SFB/TR16 and by the {\it Schweize\-ri\-scher Nationalfonds}. V.~Crede acknowledges support from the U.S. Department of Energy.


\begin{thebibliography}{}

\bibitem{Crede_Roberts} V.~Crede and W.~Roberts, Rep. Prog. Phys. \textbf{76}, 076301 (2013).

\bibitem{Klempt_Richard} E.~Klempt and J.-M.~Richard, Rev. Mod. Phys. \textbf{82}, 1095 (2010).

\bibitem{Edwards} R.~G.~Edwards \textit{et al.}, Phys. Rev. D \textbf{84}, 074508 (2011).

\bibitem{Beck_Thoma} R.~Beck and U.~Thoma, EPJ Web Conf. \textbf{134}, 02001 (2017).

\bibitem{Ireland} D.~G.~Ireland,~E.~Pasyuk and I.~Strakovsky, Prog. Part. Nucl. Phys. \textbf{111}, 103752 (2020).

\bibitem{PDG} P.~A.~Zyla  \textit{et al.}, Particle Data Group, Prog. Theor. Exp. Phys. \textbf{2020}, 083C01 (2020).

\bibitem{Assafiri} Y. Assafiri \textit{et al.}, Phys. Rev. Lett. \textbf{90}, 222001 (2003).

\bibitem{Thoma_2pi}U. Thoma \textit{et al.}, Phys. Lett. B \textbf{659}, 87 (2008).

\bibitem{Kashevarov_2pi}V. L. Kashevarov \textit{et al.}, Phys. Rev. C \textbf{85}, 064610 (2012).

\bibitem{Zehr}F. Zehr \textit{et al.}, Eur. Phys. J. A \textbf{48}, 98 (2012).

\bibitem{Oberle}M. Oberle \textit{et al.}, A2 Collaboration, Phys. Lett. B \textbf{721}, 237 (2013).

\bibitem{Dieterle}M. Dieterle \textit{et al.}, A2 Collaboration, Eur. Phys. J. A \textbf{51}, 142 (2015).

\bibitem{Sokhoyan_2pi_PLB}V. Sokhoyan \textit{et al.}, Phys. Lett. B \textbf{746}, 127 (2015).

\bibitem{Sokhoyan_2pi}V.~Sokoyan \textit{et al.}, CBELSA/TAPS Collaboration, Eur. Phys. J. A \textbf{51}, 95 (2015).

\bibitem{Thiel}A.~Thiel \textit{et al.}, CBELSA/TAPS Collaboration, Phys. Rev. Lett. \textbf{114}, 091803 (2015).

\bibitem{Ajaka}J.~ Ajaka \textit{et al.}, Phys. Rev. Lett. \textbf{100}, 052003 (2008).

\bibitem{Kashevarov_pi-eta}V.~L.~Kashevarov \textit{et al.}, Eur. Phys. J. A \textbf{42}, 141 (2009).

\bibitem{Gutz} E.~Gutz \textit{et al.}, CBELSA/TAPS Collaboration, Eur. Phys. J. A \textbf{50}, 74 (2014).

\bibitem{Kaeser} A.~K\"aser \textit{et al.}, Eur. Phys. J. A \textbf{52}, 272 (2016).

\bibitem{Sokhoyan_pi_eta} V.~Sokhoyan \textit{et al.}, A2 Collaboration, Phys. Rev. C \textbf{97}, 055212 (2018).

\bibitem{Metag_Nanova}V.~Metag M.~Nanova, EPJ Web of Conferences \textbf{199}, 020008 (2019).

\bibitem{Kuznetsov}V.~Kuznetsov \textit{et al.}, JETP Lett. \textbf{106}, 693 (2017).

\bibitem{Diakonov} D. Diakonov, V. Petrov, and M. V. Polyakov, Z. Phys. A \textbf{359}, 305 (1997).

\bibitem{Werthmuller} D. Werthm{\"u}ller \textit{et al.}, A2 Collaboration, EPJConf. \textbf{241} 01019 (2020).

\bibitem{Anisovich} A. V. Anisovich \textit{et al.}, Phys. Rev. C \textbf{95}, 035211 (2017).

\bibitem{Witthauer} L. Witthauer \textit{et al.}, A2 Collaboration, Phys. Rev. Lett. \textbf{117}, 132502 (2016).

\bibitem{Husmann_Schwille}D.~Husmann and W.~J.~Schwille, Phys. Bl. \textbf{44}, 40 (1988).

\bibitem{Hillert}W.~Hillert, Eur. Phys. J. A \textbf{28}, 139 (2006).

\bibitem{Aker}E.~Aker {\it et al.}, The Crystal Barrel Collaboration, Nucl. Instr. Meth. A \textbf{321}, 69 (1992).

\bibitem{TAPS1}R.~Novotny, IEEE Trans. Nucl. Sci. \textbf{38}, 379 (1991).

\bibitem{TAPS2}A.~R.~Gabler \textit{et al.}, Nucl. Instr. Meth. A \textbf{346}, 168 (1994).

\bibitem{Suft} G. Suft \textit{et al.}, Nucl. Instr. Meth. A \textbf{538}, 416 (2005).

\bibitem{Gottschall} M.~Gottschall \textit{et al.}, Eur. Phys. J. A \textbf{57}, 40 (2021).

\bibitem{Hartmann} J.~Hartmann \textit{et al.}, Phys. Lett. B \textbf{748}, 212 (2015).

\bibitem{Pee}H.~van Pee \textit{et al.}, Eur. Phys. J. A \textbf{31}, 61 (2007).

\bibitem{GEANT} R.~Brun \textit{et al.},  GEANT, Cern/DD/ee/84-1 (1986).

\bibitem{Klempt} E.~Klempt, A.V.~Sarantsev, and U.~Thoma, EPJ Web of Conferences \textbf{134}, 020002 (2017).

\bibitem{Kashevarov_eta-eta'} V. L.Kashevarov  \textit{et al.}, Phys. Rev. Lett. \textbf{118}, 212001 (2017).

\bibitem{Crede_eta-eta'} V.~Crede \textit{et al.}, CBELSA/TAPS Collaboration, PRC  \textbf{80}, 055202 (2009).

\bibitem{Kaeser_D}  A.~K\"aser \textit{et al.}, Phys. Lett. B. \textbf{786}, 305 (2018).

\bibitem{Doering} M.~D\"oring, E.~Oset, and D. Strottman, Phys. Rev. DC \textbf{73}, 045209 (2006).

\bibitem{Aubert} B.~Aubert \textit{et al.}, Phys. Rev. D  \textbf{31}, 112006 (2004).

\bibitem{Guo} F.~K.~Guo  \textit{et al.}, Rev. Mod. Phys. \textbf{90}, 015004 (2018).

\bibitem{COMPASS} C.~ Adolph \textit{et al.}, Phys. Rev. Lett. \textbf{115}, 082001 (2015).

\bibitem{Ketzer} M.~Mikhasenko, B.~Ketzer, and A.~Sarantsev, Phys. Rev. D  \textbf{91}, 094015 (2015).

\bibitem{Bayar} M.~Bayar \textit{et al.}, Phys. Rev. D \textbf{94}, 074039 (2016).

\bibitem{Flatte} S~.M.~Flatte, Phys. Lett. B  \textbf{63}, 224 (1976).

\bibitem{Bugg1} D.~V.~Bugg \textit{et al.}, Phys. Rev. D \textbf{50}, 4412 (1994).

\bibitem{Bugg2} D.~V.~Bugg \textit{et al.}, Phys. Rev. D \textbf{78}, 074023 (2008).

\bibitem{Celentano} A.~Celentano \textit{et al.}, Phys. Rev. C \textbf{102}, 032201 (2020).

\bibitem{Landau} L. D. Landau, Nucl. Phys. \textbf{13},  181 (1959).

\bibitem{Schmid}  C.~Schmid, Phys. Rev. \textbf{154}, 1363 (1967).

\bibitem{Aceti} F.~Aceti, L.~R.~Dai, and E.~Oset, Phys. Rev. D \textbf{94}, 096015 (2016).

\bibitem{Xie_Guo} J.-J.~Xie and F-G.~Guo, Phys. Lett. B. \textbf{774}, 108 (2017).

\bibitem{Debastiani} V.~R.~Debastiani, S.~Sakai, and E.~Oset, Eur. Phys. J. C \textbf{79}, 69 (2019).

\bibitem{Brack1} J.~T.~Brack \textit{et al.}, Phys. Rev. C \textbf{51}, 929 (1995).

\bibitem{Brack2} J.~T.~Brack \textit{et al.}, Phys. Rev. C \textbf{34}, 1771(1986).

\bibitem{Ritchie} B.~G.~Ritchie \textit{et al.}, Phys. Lett. B \textbf{125}, 128 (1983).

\bibitem{SAID} R.~L.~Workman \textit{et al.}, Phys. Rev. C \textbf{86}, 035202 (2012).

\bibitem{Khemchandani} K.~P.~Khemchandani \textit{et al.}, Phys. Rev. D \textbf{103}, 016015 (2021).

\bibitem{Meissner} U.-G.~Mei\ss ner, Symmetry \textbf{12}, 981 (2020).








\end{thebibliography}
\end{document}